\documentclass[%
 reprint,
 amsmath,amssymb,
 aps,
]{revtex4-2}

\usepackage{multirow}
\usepackage{graphicx}
\usepackage{dcolumn}
\usepackage{bm}
\usepackage{hyperref}
\usepackage{CJKutf8} 
\usepackage[dvipsnames]{xcolor}

\begin{document}

\preprint{APS/123-QED}

\title{A bounded-confidence model of opinion dynamics with heterogeneous node-activity levels}

\author{Grace J. Li}
\affiliation{Department of Mathematics, University of California, Los Angeles, California 90095, USA}
\author{Mason A. Porter}
 \email{mason@math.ucla.edu}
\affiliation{Department of Mathematics, University of California, Los Angeles, California 90095, USA}
\affiliation{Santa Fe Institute, Santa Fe, New Mexico 87501, USA}

\date{\today}

\begin{abstract}
Agent-based models of opinion dynamics allow one to examine the spread of opinions between entities and to study phenomena such as consensus, polarization, and fragmentation. By studying a model of opinion dynamics on a social network, one can explore the effects of network structure on these phenomena. In social networks, some individuals share their ideas and opinions more frequently than others. These disparities can arise from heterogeneous sociabilities, heterogeneous activity levels, different prevalences to share opinions when engaging in a social-media platform, or something else. 
To examine the impact of such heterogeneities on opinion dynamics, we generalize the Deffuant--Weisbuch (DW) bounded-confidence model (BCM) of opinion dynamics by incorporating node weights. The node weights allow us to model agents with different probabilities of interacting. Using numerical simulations, we systematically investigate (using a variety of network structures and node-weight distributions) the effects of node weights, which we assign uniformly at random to the nodes. We demonstrate that introducing heterogeneous node weights results in longer convergence times and more opinion fragmentation than in a baseline DW model. The node weights in our BCM allow one to consider a variety of sociological scenarios in which agents have heterogeneous probabilities of interacting with other agents.
\end{abstract}


\maketitle


\section{Introduction} \label{sec:intro}
Humans are connected in numerous ways, and our many types of interactions with each other influence what we believe and how we act. 
To model how opinions spread between people or other agents, researchers across many disciplines have developed a variety of models of opinion dynamics \cite{castellano2009, sirbu_book2017, sune-yy2018, noorazar-et-al2020, noorazar2020, peralta2022, galesic2021}.
However, in part because of the difficulty of gathering empirical data on opinions, much of the research on opinion dynamics has focused on theory and model development, with little empirical validation \cite{castellano2009, galesic2021, peralta2022, vazquez2022}.
Some researchers have examined how human opinions change in controlled experimental settings with questionnaires~\cite{chacoma2015,vandeKerckhove2016, takacs2016}, 
and others have examined empirical opinion dynamics using data from social-media platforms~\cite{monti2020, kozitsin2022_real_data, kozitsin2023_micro_level}.
One of the many difficulties in empirically validating models of opinion dynamics is the potential sensitivity of model outcomes to measurement errors of real-life opinion values \cite{carpentras2021}.
See M\"as~\cite{mas2019} for a discussion of some of the challenges of validating models in the social sciences.
Even with the difficulty of validating models of opinion dynamics, it is valuable to formulate and study such models. Developing mechanistic models forces researchers to clearly define assumptions, variables, and the relationships between variables; 
such models provide frameworks to explore and generate testable hypotheses about complex social phenomena \cite{vazquez2022, holme2015}.

In an agent-based model (ABM) of opinion dynamics, each agent is endowed with an opinion and an underlying network structure governs which agents can interact with each other. 
We assume that all interactions are dyadic (i.e., between exactly two agents), and we suppose that the agent opinions take continuous values in a closed interval on the real line~\footnote{
There are also many models of opinion dynamics with discrete-valued opinions and/or polyadic interactions between agents~\cite{noorazar-et-al2020, sirbu_book2017, battiston2020networks}.}.
This interval represents a continuous spectrum of views about something, such an ideology or the strength of support for a political candidate. 
At each discrete time step of an ABM of opinion dynamics, one selects which agents interact and then use an update rule to determine if and how their opinions change. 
Bounded-confidence models (BCMs) are a popular class of models with continuous-valued opinions \cite{noorazar-et-al2020}.
In a BCM, interacting agents influence each other only when their opinions are sufficiently similar. This mechanism is reminiscent of
the psychological idea of selective exposure, which asserts that people tend to seek information or conversations that support their existing views and avoid those that challenge their views~\cite{selective_exposure_def}. 
Under this assumption, an agent's views are influenced directly only by agents with sufficiently similar views. For example, social-media platforms include polarizing posts, but individuals can choose whether or not to engage with such content; they do not adopt the views of everything in their social-media feeds.

The two most popular BCMs are the Hegselmann--Krause (HK) model \cite{hegselmann_krause2002} and the Deffuant--Weisbuch (DW) model \cite{deffuant2000}. 
At each time step, the HK model has synchronous updates of node opinions, whereas the DW model has asynchronous opinion updates, with a single pair of agents (i.e., a dyad) interacting and potentially updating their opinions at each time. An asynchronous mechanism is consistent with empirical studies, which suggest that individuals in social networks have different activity times and frequencies~\cite{alizadeh2015}.
In the present paper, we generalize the DW model to incorporate heterogeneous node-activity levels.
Although the DW model has been generalized in many ways \cite{noorazar2020},
few studies have modified the procedure to select which agents interact in a time step.
The ones that have modified this procedure
(see, e.g., Refs.~\cite{alizadeh2015, zhang2018, sirbu2019, pansanella2022}) have focused on specific scenarios, rather than on investigating the effects of introducing heterogeneities into agent-selection probabilities.

Before we describe previous extensions of the DW model that incorporate heterogeneities in agent selection, we first discuss other generalizations of the model.
The DW model was first studied on complete graphs \cite{deffuant2000}. 
To explore the effects of network structure on DW dynamics, many researchers subsequently simulated DW models on time-independent graphs \cite{meng2018}.
Researchers have also examined DW models on hypergraphs \cite{hickok2022} and coevolving networks \cite{unchitta2021}.
Additionally, many studies have extended the DW model to consider different initial conditions and/or BCM parameters.
Some studies have considered initial node opinions that arise from nonuniform distributions \cite{jacobmeier2006, carro2013, sobkowicz2015,hickok2022}, yielding initial conditions that are different from those in the standard DW model. 
Other investigations have incorporated heterogeneous confidence bounds or heterogeneous opinion compromises 
\cite{dw2002, deffuant-amblard2002, lorenz2008, kou2012, zhang2014, huang2018, sobkowicz2015, chen2020}.
Such generalizations affect the opinion updates of interacting agents.

In the standard DW model, one selects pairs of agents to interact uniformly at random, but social interactions are not uniform in real life.
Few studies of the DW model have modified the selection procedure that determines which agents interact with each other; see, e.g.,~\cite{alizadeh2015, zhang2018, sirbu2019, pansanella2022}. 
When selecting agents in a way that is not uniformly at random, one can think of the agents as having different activity levels that encode their interaction frequencies. (In a given time interval, we expect these agents to have different numbers of interactions.)
The idea of heterogeneous node-activity levels plays an important role in activity-driven models of temporal networks \cite{perra2012}.
There have also been studies of activity-driven models of opinion dynamics. Li et al.~\cite{li2017} developed an activity-driven model of opinion dynamics using networks with fixed nodes with assigned activity rates (i.e., assigned activation probabilities).
At each time step of their model, one removes all existing edges and the active agents randomly form a fixed number of connections. All agents then evaluate the mean opinions of their neighbors to determine if and how to update their own opinions \cite{li2017}.
Baronchelli et al.~\cite{baronchelli2011} studied a voter model with heterogeneous edge weights, which one can interpret as encoding heterogeneous edge activities.

Some researchers have generalized the DW model to incorporate heterogeneous agent selection. Alizadeh and Cioffi-Revilla \cite{alizadeh2015} studied a modified DW model that incorporates a repulsion mechanism (which was proposed initially by Huet et al.~\cite{huet2008}) in which interacting agents with opinions that differ by more than a cognitive-dissonance threshold move farther away from each other in the space of opinions. They used two-dimensional (2D) vector-valued opinions and placed their nodes on complete graphs. To model agents with different activity levels, Alizadeh and Cioffi-Revilla \cite{alizadeh2015} implemented a Poisson node-selection probability, which one can interpret as independent internal ``clocks'' that determine agent activation. In comparison to selecting agent pairs uniformly at random (as in the standard DW model) the Poisson node-selection probability can either lessen or promote the spread of extremist opinions, depending on which opinions are more prevalent in more-active agents.

Zhang et al.~\cite{zhang2018} examined a modified DW model with asymmetric updates on activity-driven networks. In their model, each node has a fixed activity potential, which one assigns uniformly at random from a distribution of activity potentials. The activity potential of an agent is its probability to activate. At each discrete time step, each active agent $i$ randomly either (1) creates a message (e.g., a social-media post) or (2) boosts a message that was created by a neighboring agent $j$. If agent $i$ boosts a message from agent $j$, then $i$ updates its opinion using the standard DW update mechanism.
Zhang et al.~\cite{zhang2018} simulated their model on a social network from Tencent Weibo
(\begin{CJK*}{UTF8}{gbsn}腾讯微博\end{CJK*})
and found that the distribution of activity potentials influences the location of the transition between opinion consensus and fragmentation.
The node weights in our BCM are similar in spirit to the activity potentials of Zhang et al.~\cite{zhang2018}; they can encode the social activity levels of individuals, such as their frequencies of posting or commenting on social media. 
However, the way that we incorporate node weights in our BCM differs fundamentally from Ref.~\cite{zhang2018}.
We consider a time-independent network $G$, and we select a single pair of neighboring agents to interact at each time step. We first randomly select one agent with a probability that is proportional to its node weight, and then we randomly select a second neighboring agent with a probability that depends on its node weight.
The two selected agents then update their opinions using the DW update mechanism.

Heterogeneities in which interactions occur in a social network arise not only because some individuals are more likely to have interactions, but also because some pairs of individuals are more likely to interact than other pairs \cite{baronchelli2011}.
The curation of content in social-media feeds is affected by homophily, which is the idea that individuals have a tendency to connect with others that are similar to themselves (e.g., perhaps they have similar ideas or beliefs) \cite{mcpherson2001}.
Social-media feeds tend to show content to users that closely matches their profiles and past activities \cite{spohr2017}. 
To examine the effect of such algorithmic bias on opinion dynamics, S\^{i}rbu et al.~\cite{sirbu2019} studied a modified DW model that includes a homophily-promoting activation mechanism. 
At each time step, one agent is selected uniformly at random, and then one of its neighbors is selected with a probability that depends on the magnitude of the opinion difference between that neighbor and the first agent. The simulations by S\^{i}rbu et al. of their model on complete graphs suggest that more algorithmic bias yields slower convergence times and more opinion fragmentation \cite{sirbu2019}. 
Pansanella et al.~\cite{pansanella2022} applied the same algorithmic-bias model to a variety of network topologies (specifically, Erd\H{o}s--R\'{e}nyi, Barab\'{a}si--Albert, and Lancichinetti--Fortunato--Radicchi (LFR) graphs), and they found similar trends as S\^{i}rbu et al. did on complete graphs.

From the investigations in Refs.~\cite{alizadeh2015, zhang2018, sirbu2019, pansanella2022}, 
we know that incorporating heterogeneous node-selection probabilities into a DW model can influence opinion dynamics.
Each of these papers examined a specific implementation of heterogeneous agent selection; we are not aware of any systematic investigations of the effects of heterogeneous agent selection on opinion dynamics in asynchronous BCMs.
In the present paper, we propose a novel BCM with heterogeneous agent-selection probabilities, which we implement using node weights.
In general terms, we are studying a dynamical process on node-weighted networks.
We use node weights to model agents with different probabilities of interacting. These probabilities can encode heterogeneities in individual behavior, such as in sociability or activity levels.
We conduct a methodical investigation of the effects of incorporating heterogeneous node weights, which we draw from various distributions, into our generalization of the DW model. 
We examine these effects on a variety of types of networks. 
In our study, we consider fixed node weights that we assign in a way that disregards network structure and node opinions. However, one can readily adapt the node weights in our BCM to consider a variety of sociological scenarios in which nodes have heterogeneous selection probabilities.
We find that introducing heterogeneous node weights into our node-weighted BCM results in longer convergence times and more opinion fragmentation than selecting nodes uniformly at random.
Our results illustrate that it is important to consider the influence of assigning node-selection probabilities uniformly at random in 
models with heterogeneous node selection before drawing conclusions about more specific mechanisms such as algorithmic bias \cite{sirbu2019}.
More generally, our model illustrates the importance and utility of incorporating node weights into network analysis and dynamics.

Our paper proceeds as follows. In Sec.~\ref{sec:model}, we describe the standard DW model and present our generalized DW model with node weights to incorporate heterogeneous agent-selection probabilities.
In Sec.~\ref{sec:methods}, we discuss the setup of our simulations, the networks and node-weight distributions that we examine, and the quantities that we compute to characterize the behavior of our model.
In Sec.~\ref{sec:results}, we discuss the results of our numerical simulations of our BCM. 
In Sec.~\ref{sec:discussion}, we summarize our results and discuss their implications, present some ideas for future work, and highlight the importance of studying networks with node weights.
Our code is available at \url{https://gitlab.com/graceli1/NodeWeightDW}.


\section{Our model} \label{sec:model} 
In this section, we first discuss the Deffuant--Weisbuch (DW) \cite{deffuant2000} bounded-confidence model (BCM) of opinion dynamics, and we then introduce our BCM with heterogeneous node-selection probabilities.


\subsection{The standard Deffuant--Weisbuch (DW) BCM} \label{sec:DW}
The DW model was introduced over two decades ago \cite{deffuant2000}, and this model and its extensions have been studied extensively since then \cite{noorazar-et-al2020, noorazar2020}. The DW model was examined originally on complete graphs and encoded agent 
opinions as scalar values in a closed interval on the real line. 
Deffuant et al.~\cite{deffuant2000} let each agent  
have an opinion in $[0,1]$, and we follow this convention. 
The standard DW model has two parameters. The ``confidence bound'' $c \in [0,1]$ is a thresholding parameter; 
when two agents interact, they compromise their opinions by some amount if and only if their opinions differ by less than $c$.
The ``compromise parameter'' $m \in (0, 0.5]$ (which is also sometimes called a convergence parameter \cite{deffuant2000} or a cautiousness parameter \cite{meng2018}) parametrizes the amount that an agent changes its opinion to compromise with the opinion of an agent with whom it interacts.

In the standard DW model, the opinions of the agents update asynchronously. We endow each agent with an initial opinion. At each discrete time, one uniformly randomly selects a pair of agents to interact.
At time $t$, suppose that we pick agents $i$ and $j$, whose associated opinions are $x_i$ and $x_j$, respectively. Agents $i$ and $j$ update their opinions through the following equations:
\begin{align} \label{eq:DW}
\begin{split}
    x_i(t+1) &= 
    \begin{cases}
        x_i(t) + m \Delta_{ji}\,, & \text{if } |\Delta_{ij}(t)| < c \\
        x_i(t)\,, & \text{otherwise} \,,
    \end{cases} \\
    x_j(t+1) &= 
    \begin{cases}
        x_j(t) + m \Delta_{ij}\,, & \text{if } |\Delta_{ij}(t)| < c \\
        x_j(t)\,, & \text{otherwise}\,,
    \end{cases}
\end{split}
\end{align}
where $\Delta_{ij}(t) = x_i(t) - x_j(t)$. When $|\Delta_{ij}(t)| < c$, we say that agents $i$ and $j$ are ``receptive'' to each other at time $t$. When $|\Delta_{ij}(t)| \geq c$, we say that agents $i$ and $j$ are ``unreceptive'' to each other.

When one extends the DW model to consider an underlying network of agents \cite{weisbuch2001}, only adjacent agents are allowed to interact. 
Consider an undirected network $G = (V, E)$, where $V$ is the set of nodes and $E$ is the set of edges between them. Let $N = |V|$ denote the size of the network (i.e., the number of nodes of the network).
Each node of a network represents an agent, and each edge between two agents encodes a social or communication tie between them. 
At each discrete time, one selects an edge of a given network uniformly at random and the two agents that are attached to that edge interact with each other; they update their opinions following Eq.~\eqref{eq:DW}.
For the DW model, an alternative to an edge-based approach of randomly selecting an interacting edge is to take a node-based approach to determine the agents that interact.
(See Ref.~\cite{kureh2020} for a discussion of node-based updates versus edge-based updates in the context of voter models.)
In a node-based approach, one first randomly selects one node and then randomly selects a second node from its neighbors. 
To capture the fact that some agents have more frequent interactions (such as from greater sociability or a stronger desire to share their opinions) than others, we implement a node-based agent-selection procedure in our study. 

The choice between edge-based and node-based agent selection can have substantial effects on the dynamics of voter models of opinion dynamics \cite{kureh2020}, and we expect that this is also true for other types of opinion-dynamics models. We are not aware of a comparison of edge-based and node-based agent selection in asynchronous BCMs (and, in particular, in DW models), and it seems both interesting and relevant to explore this issue. Most past research on the DW model has considered edge-based selection \cite{noorazar2020}. However, Refs.~\cite{alizadeh2015, sirbu2019, pansanella2022} used a node-based selection procedure to model heterogeneous activities of agents.


\subsection{A BCM with heterogeneous node-selection probabilities} \label{sec:BCM} 
We now introduce our BCM with heterogeneous node-selection probabilities.
Consider an undirected network $G = (V, E)$. As in the standard DW model, suppose that each agent $i$ has a time-dependent opinion $x_i(t)$.
In our BCM, each agent also has a fixed node weight $w_i$ that encodes sociability, how frequently it engages in conversations, or simply the desire to share its opinions. 
One can think of a node's weight as a quantification of how frequently it talks to its friends or posts on social media.
By incorporating network structure, the standard DW model can include agents with different numbers of friends (or other social connections).
However, selecting interacting node pairs uniformly at random is unable to capture the heterogeneous interaction frequencies of individuals.
By introducing node weights, we encode such heterogeneity
and then examine how it affects opinion dynamics in a BCM.
Although we employ fixed node weights, one can adapt our model to include time-dependent node weights, such as through purposeful strategies (such as posting on social media more frequently as one's opinions become more extreme).

In our node-weighted BCM, at each discrete time, we first select an agent $i$ with a probability that is proportional to its weight. Agent $i$ then interacts with a neighbor $j$, which we select with a probability that is equal to its weight divided by the sum of the weights of $i$'s neighbors.
That is, the probabilities of first selecting agent $i$ and then selecting agent $j$ are
\begin{align} \label{eq:node-probablity}
    P_1(i) = \frac{w_i}{\sum\limits_{k = 1}^N w_k} \,,  \quad
    P_2(j|i) = \frac{w_j}{\sum\limits_{k \in \mathcal{N}(i)} w_k}\,, 
\end{align}
where $\mathcal{N}(i)$ denotes the neighborhood (i.e., the set of neighbors) of node $i$. Once we select the pair of interacting agents, we update their opinions following the DW opinion update rule in Eq.~\ref{eq:DW}.

Our BCM incorporates heterogeneous node-selection probabilities
with node weights that model phenomena such as the heterogeneous sociability of individuals. 
One can also study heterogeneous selection probabilities
of pairwise (i.e., dyadic) interactions, instead of focusing on the probabilities of selecting individuals.
For instance, an individual may discuss their ideological views with a close friend more frequently than with a work colleague.
One can use edge weights to determine the probabilities of selecting the dyadic interactions in a BCM.
At each discrete time, one can select an edge with a probability that is proportional to its weight. We do not examine edge-based heterogeneous selection probabilities in the present paper, but it is worth exploring in BCMs.

\section{Methods and simulation details} \label{sec:methods} 
In this section, we discuss the network structures and node-weight distributions that we consider, the setup of our numerical simulations, and the quantities that we compute to characterize the results of our simulations.


\subsection{Network structures} \label{sec:nets}
We now describe the details of the networks on which we simulate our node-weighted BCM. We summarize these networks in Table~\ref{tab:networks}.

We first simulate our BCM on complete graphs as a baseline scenario that will allow us to examine how incorporating heterogeneous node-selection probabilities affects opinion dynamics.
Although DW models were introduced more than 20 years ago, complete graphs are still the most common type of network on which to study them \cite{noorazar-et-al2020}. 
To examine finite-size effects from our networks, we consider complete graphs with sizes $N \in \{10, 20, 30, 45, 65, 100, 150, 200, 300, \ldots, 1000\}$.
For all other synthetic networks, we consider networks with $N = 500$ nodes.

\newcommand{\leftcol}{1.5in}
\newcommand{\midcol}{3.55in}
\newcommand{\rightcol}{1.5in}

\begin{table*}[htb]
\centering
\caption{\label{tab:networks} The networks on which we simulate our node-weighted BCM.}
\begin{ruledtabular}
\def\arraystretch{1.1} 
\begin{tabular}{m{\leftcol} m{\midcol} m{\rightcol}}
\textbf{Network} & \textbf{Description} & \textbf{Parameters} \\\hline
$C(N)$   
& \begin{tabular}{m{\midcol}} 
Complete graph with $N$ nodes \end{tabular}
& \begin{tabular}[c]{m{\rightcol}}
$N \in \{10, 20, 30, 45, 65, 100,$\\
$150, 200, 300 \dots , 1000\}$ 
\end{tabular}   
\\\hline
$G(N,p)$             
& \begin{tabular}{m{\midcol}} Erd\H{o}s--R\'{e}nyi (ER) random-graph model with $N$ nodes and homogeneous, independent edge probability $p$ \end{tabular}
& \begin{tabular}[c]{m{\rightcol}} 
$\, p \in \{0.1, 0.3, 0.5, 0.7\}$ \end{tabular} 
\\\hline
Two-Community SBM\footnotemark[1]  
& \begin{tabular}{m{\midcol}} Stochastic block model with 2 $\times$ 2 blocks. Edges between nodes in the same set (A or B) exist
with a larger probability than edges between nodes in different sets;
the block probabilities satisfy $P_{BB} > P_{AA} > P_{AB}$. \end{tabular}
& \begin{tabular}[c]{m{\rightcol}}$P_{AA} = 49.9/374$ \\ 
$P_{BB} = 49.9/124$ \\ $P_{AB} = 1/500$ \end{tabular}
\\\hline
Core--Periphery SBM\footnotemark[1] 
& \begin{tabular}{m{\midcol}} Stochastic block model with 2 $\times$ 2 blocks. Set A is a set of core nodes and set B is a set of peripheral nodes. The block probabilities satisfy $P_{AA} > P_{AB} > P_{BB}$. \end{tabular}  
& \begin{tabular}[c]{m{\rightcol}}$P_{AA} = 147.9/374$ \\ 
$P_{BB} = 1/174$ \\ $P_{AB} = 1/25$ \end{tabular}          
\\\hline
Caltech Network
& \begin{tabular}{m{\midcol}} The largest connected component of the Facebook friendship network at Caltech on one day in fall 2005. This network, which is part of the {\sc Facebook100} data set \cite{red2011, traud2012}, has 762 nodes and 16,651 edges. \end{tabular}
&      
\end{tabular}
\end{ruledtabular}
\footnotetext[1]{Our SBM networks have $N = 500$ nodes. We partition an SBM network into two sets of nodes; set A has 75\% of the nodes, and set B has 25\% of the nodes.}
\end{table*}

We consider synthetic networks that we generate using
the $G(N, p)$ Erd\H{o}s--R\'{e}nyi (ER) random-graph model, where $p$ is the homogeneous, independent probability of an edge between each pair of nodes \cite{newman2018}. When $p = 1$, this yields a complete graph. We examine $G(500, p)$ graphs with $p \in \{0.1, 0.3, 0.5, 0.7\}$.

To determine how a network with an underlying block structure affects the dynamics of our node-weighted BCM, we consider stochastic-block-model (SBM) networks \cite{newman2018} with $2 \times 2$ blocks, where each block consists of an ER graph.
Inspired by the choices of Kureh and Porter \cite{kureh2020}, we consider two types of SBM networks. The first has a two-community structure, in which there is a larger probability of edges within a community than between communities. 
The second SBM has a core--periphery structure, in which there is a set of core nodes with a large probability of edges within the set,
a set of peripheral nodes with a small probability of edges within the set, and edges exist between core nodes and peripheral nodes with an intermediate probability.
To construct our $2 \times 2$ SBMs, we partition a network into two sets of nodes; set A has 375 nodes (i.e., 75\% of the network) and set B has 125 nodes (i.e., 25\% of the network). We define a symmetric edge-probability matrix
\begin{equation}
    P = \begin{bmatrix}
    P_{AA} & P_{AB} \\ P_{AB} & P_{BB}
    \end{bmatrix} \,,
\end{equation}
where $P_{AA}$ and $P_{BB}$ are the probabilities that an edge exists between two nodes in set A and set B, respectively, and $P_{AB}$ is the probability that an edge exists between a node in set A and a node in set B.

In a two-community SBM, the probabilities $P_{AA}$ and $P_{BB}$ are larger than $P_{AB}$, so edges between nodes in the same community exist with a larger probability than edges between nodes in different communities.
For our two-community SBM, we choose $P_{AA}$ and $P_{BB}$ so that the expected mean degree matches that of the $G(500, 0.1)$ ER model if we only consider edges within set A or edges within set B. A network from the $G(N, p)$ model has an expected mean degree of $p(N-1)$ \cite{newman2018}, so we want the two communities of these SBM networks to have an expected mean degree of $49.9 = 0.1 \times 499$. We thus use the edge probabilities $P_{AA} = 49.9/374$ and $P_{BB} = 49.9/124$. To ensure that there are few edges between the sets A and B, we choose $P_{AB} = 1/500$.

We want our core--periphery SBM with core set A and periphery set B to satisfy $P_{AA} > P_{AB} > P_{BB}$.
We chose $P_{AA}$ so that the expected mean degree matches that of the $G(500, 0.3)$ model (i.e., it is 147.9) if we only consider edges within the set A. We thus choose the edge probability $P_{AA} = 147.9/374$. To satisfy $P_{AA} > P_{AB} > P_{BB}$, we choose $P_{AB} = 1/25$ and $P_{BB} = 1/174$.

Finally, we investigate our node-weighted BCM on a real social network from Facebook friendship data. We use the Caltech network from the {\sc Facebook100} data set; its nodes encode individuals at Caltech, and its edges encode Facebook ``friendships'' between them on one day in fall 2005 \cite{red2011, traud2012}. We only consider the network's largest connected component, which has 762 nodes and 16,651 edges.


\subsection{Node-weight distributions} \label{sec:weights}
In Table~\ref{tab:distributions}, we give the parameters and probability density functions of the node-weight distributions that we examine in our BCM. In this subsection, we discuss our choices of distributions.

\begin{table*}
\centering
\caption{\label{tab:distributions} Names and specifications of our distributions of node weights. We show both the general mathematical expressions for the means and the specific values of the means for our parameter values.
For the Pareto distributions, the distribution means in the table are approximate. For all other distributions, the means are exact.}
\def\arraystretch{1.2} 
\begin{ruledtabular}
\begin{tabular}{lclccl}
\textbf{Distribution} & 
\textbf{\begin{tabular}[c]{@{}c@{}}Probability density \\ function\end{tabular}} & 
\textbf{Parameter values}     & \textbf{Domain}           & \multicolumn{2}{c}{\textbf{Mean}}  \\ \colrule
Constant & $\delta (x - 1)$  & N/A & $\{1\}$   & 1 & 1 \\ \colrule
Pareto-80-10 &
  \multirow{3}{*}{$\dfrac{\alpha}{x^{\alpha+1}}$} &
  $\alpha = \log_{4.5}(10)$ &
  \multirow{3}{*}{$[1,\infty)$} &
  \multirow{3}{*}{$\dfrac{\alpha}{\alpha-1}$} &
  2.8836 \\
Pareto-80-20 &    & $\alpha = \log_{4}(5)$ &   &   & 7.2126 \\
Pareto-90-10 &    & $\alpha = \log_9 (10)$ &   &   & 21.8543  \\ \colrule
Exp-80-10 &
  \multirow{3}{*}{$\frac{1}{\beta} \exp\left(\frac{-(x-1)}{\beta}\right)$} &
  $\beta = 1.8836$ &
  \multirow{3}{*}{$[1,\infty)$} &
  \multirow{3}{*}{$\beta + 1$} &
  2.8836 \\
Exp-80-20     &  & $\beta = 6.2125$  & & & 7.2125        \\
Exp-90-10     &  & $\beta = 20.8543$ & & & 21.8543       \\  \colrule
Unif-80-10              & 
    \multirow{3}{*}{$\dfrac{1}{b-1}$}      & 
    $b = 4.7672$                & 
    \multirow{3}{*}{$[1, b]$} & 
    \multirow{3}{*}{$\dfrac{1}{2} (1 + b)$} & 
    2.8836           \\
Unif-80-80         &  & $b = 13.425$      & & & 7.2125        \\
Unif-90-10         &  & $b = 42.7086$     & & & 21.8543       \\
\end{tabular}
\end{ruledtabular}
\end{table*}

To study the effects of incorporating node weights in our BCM, we compare our model to a baseline DW model.
To ensure a fair comparison, we implement a baseline DW model that selects interacting agents uniformly at random using a node-based selection process.
As we discussed in Sec.~\ref{sec:intro}, it is much more common to employ an edge-based selection process.
We refer to the case in which all nodes weights are equal to $1$ (that is, $w_i = 1$ for all nodes $i$) as the ``constant weight distribution''. The constant weight distribution (and any other situation in which all node weights equal the same positive number) results in a uniformly random selection of nodes for interaction. 
This is what call the ``baseline DW model''; we compare our DW models with heterogeneous node weights to this baseline model.
We reserve the term ``standard DW model'' for the DW model with uniformly random edge-based selection of agents.

The node weights in our BCM encode heterogeneities in interaction frequencies, such as when posting content online.
The majority of online content arises from a minority of user accounts~\cite{guo2009}.
A ``90-9-1 rule'' has been proposed for such participation inequality. In this rule of thumb, about 1\% of the individuals in online discussions (e.g., on social-media platforms) account for most contributions, about 9\% of the individuals contribute on occasion, and the remaining 90\% of the individuals are present online (e.g., they consume content) but do not contribute to it \cite{90-9-1}.
Participation inequality has been documented in a variety of situations, including the numbers of posts on digital-health social networks \cite{vanMierlo2014}, 
posts on internet support groups \cite{carron2014}, and contributions to open-source software-development platforms \cite{gasparini2020}.
Inequality in user activity has also been examined on Twitter \cite{antelmi2019}, and Xiong and Liu~\cite{xiong2014} used a power-law distribution to model the number of tweets about different topics.
A few years ago, a survey by the Pew Research Center found that about 10\% of the accounts of adult Twitter users in the United States generate about 80\% of the tweets of such accounts \cite{twitter_stats}.

One can interpret the node weights in our BCM as encoding the participation of individuals in the form of contributing content to a social-media platform.
We model online participation inequality by using a Pareto distribution for the node weights. This choice of distribution is convenient because of its simple power-law form. 
It has also been used to model inequality in a variety of other contexts, including distributions of wealth, word frequencies, website visits, and numbers of paper citations \cite{newman2005}.
When representing social-media interactions, we only care about accounts that make posts or comments; we ignore inactive accounts. Therefore, we impose a minimum node weight in our model.
We use the Pareto type-I distribution, which is defined on $[1, \infty)$, so each node has a minimum weight of $1$. This positive minimum weight yields a reasonable convergence time for the simulations of our BCM. 
Nodes with weights close to $0$ would have very small probabilities of interacting, and allowing such weights would prolong simulations.

Let Pareto-X-Y denote the continuous Pareto distribution in which (in theory) X\% of the total node weight is distributed among Y\% of the nodes. 
In practice, once we determine the $N$ node weights for our simulations from a Pareteo node-weight distribution, it is not true that precisely X\% of the total weight is held by Y\% of the $N$ nodes. 
Inspired by the results of the aforementioned Pew Research Center survey of Twitter users \cite{twitter_stats}, we first consider a Pareto-80-10 distribution, in which we expect 80\% of the total weight to be distributed among 10\% of nodes.
The Pareto principle (which is also known as the ``80-20 rule'') is a popular rule of thumb that suggests that 20\% of individuals have 80\% of the available wealth \cite{newman2005}. Accordingly, we also consider a Pareto-80-20 distribution. Finally, as an example of a node-weight distribution with a more extreme inequality, we also consider a Pareto-90-10 distribution.

We also examine uniform and exponential distributions of node weights. To match the domain of our Pareto distributions, we shift the uniform and exponential distributions so that their minimum node weight is also $1$. 
We also choose their parameters to approximate the means of our Pareto distributions. 
We use Exp-X-Y and Unif-X-Y as shorthand notation to denote exponential and uniform distributions, respectively, with means that match that of the Pareto-X-Y distribution to four decimal places (see Table~\ref{tab:distributions}). 
When we examine the results of our numerical simulations, we want to compare distributions with similar means. 
We use the phrase ``80-20 distributions'' to refer to the Pareto-80-20, Exp-80-20, and Unif-80-20 distributions. We analogously use the phrases ``80-10 distributions'' and ``90-10 distributions.''
In total, we examine three different families of distributions (Pareto, exponential, and uniform) with tails of different heaviness.
In Table~\ref{tab:distributions}, we show the details of the probability density functions and the parameters of our node-weight distributions.


\subsection{Simulation specifications} \label{sec:sims} 
In our node-weighted BCM, agents have opinions in the one-dimensional (1D) opinion space $[0,1]$. Accordingly, we examine values of the confidence bound $c \in (0, 1)$
\footnote{The extreme case $c = 0$ is degenerate (because no agents update their opinions), and the case $c = 1$ allows all adjacent agents to interact with each other. We are not interested in examining these cases.}.
We examine values of the compromise parameter $m \in (0, 0.5]$, which is the typically studied range for the DW model \cite{noorazar-et-al2020, meng2018}. 
When $m = 0.5$, two interacting agents that influence each other fully compromise and average their opinions. 
When $m < 0.5$, the two agents move towards each other's opinions, but they do not change their opinions to the mean (i.e., they do not fully compromise).

In our node-weighted BCM, the generation of graphs in a random-graph ensemble, the sets of node weights, the sets of initial opinions, and the selection of pairs of agents to interact at each time step are all stochastic. We use Monte Carlo simulations to reduce these sources of noise in our simulation results.
For each of our random-graph models (i.e., the ER and SBM graphs), we generate 5 graphs.
For each graph and each node-weight distribution, we randomly generate 10 sets of node weights. For each set of node weights, we generate 10 sets of initial opinions that are distributed uniformly at random. 
In total, we consider 100 distinct sets of initial opinions and node weights for the Monte Carlo simulations of each individual graph.
When we compare simulations from different distributions of node weights in the same individual graph, we reuse the same 100 sets of initial opinions.

In theory, the standard DW model and our node-weighted DW model can take infinitely long to approach a steady state.
We define an ``opinion cluster'' $S_r$ to be a maximal connected set of agents in which the pairwise differences in opinions are all strictly less than the confidence bound $c$; adding any other agent to $S_r$ will yield at least one pair of adjacent agents with an opinion difference of at least $c$.
Equivalently, for each graph $G$, we define the ``effective-receptivity network'' $G_{\mathrm{eff}}(t) = (V, E_{\mathrm{eff}}(t))$ as the time-dependent subgraph of it with edges only between pairs of nodes that are receptive to each others' opinions.
That is,
\begin{equation}
    E_{\mathrm{eff}}(t) = \{ (i,j) \in E : |x_i(t) - x_j(t)| < c \} \,.
\end{equation}
The opinion clusters are the connected components of the effective-receptivity network $G_{\mathrm{eff}}(t)$.
If two opinion clusters $S_1$ and $S_2$ are separated by a distance of at least $c$ 
(i.e., $|x_i - x_j| \geq c$ for all $i \in S_1$ and $j \in S_2$) 
at some time $\tilde{T}$, then (because $c$ is fixed) no agents from $S_1$ can influence the opinion of an agent in $S_2$ (and vice versa) for all $t \geq \tilde{T}$. Therefore, in finite time, we observe the formation of steady-state clusters of distinct opinions.
Inspired by Meng et al.~\cite{meng2018}, we specify that one of our simulations has ``converged'' if all opinion clusters are separated from each other by a distance of at least $c$ and each opinion cluster has an opinion spread that is less than a tolerance of $0.02$. That is, for each cluster $S_r$, we have that $\max_{i, j \in S_r} |x_i - x_j| < 0.02$.
We use $T$ to denote the convergence time in our simulations; the connected components of $G_{\mathrm{eff}}(T)$ are the steady-state opinion clusters.

It is computationally expensive to numerically simulate a DW model. Additionally, as we will show in Sec.~\ref{sec:results}, our node-weighted DW model with heterogeneous node weights often converges to a steady state even more slowly than the baseline DW model.
To reduce the computational burden of checking for convergence, we do not check for it at each time step and we compute the convergence time to three significant figures. 
To guarantee that each simulation stops in a reasonable amount of time, we set a bailout time of $10^9$ time steps. 
In our simulations, the convergence time is always shorter than the bailout time. We thus report the results of our simulations as steady-state results.


\subsection{Quantifying opinion consensus and fragmentation} \label{sec:quantities}
In our numerical simulations, we investigate which situations yield consensus (specifically, they result in one ``major'' opinion cluster, which will discuss shortly) at steady state and which situations yield opinion fragmentation (when there are at least two distinct major clusters) at steady state.
\footnote{Some researchers use the term ``polarization'' to refer to the presence of exactly two opinion clusters (or to exactly two major opinion clusters) and ``fragmentation'' to refer to the presence of three or more opinion clusters (or to three or more major opinion clusters) \cite{hegselmann_krause2002, bramson2016}. However, because we are interested in distinguishing between consensus states and any state that is not a consensus, we use the term ``fragmentation'' for any state with at least two major opinion clusters. We then quantify the extent of opinion fragmentation.}
We are also interested in how long it takes to determine the steady-state behavior of a simulation and in quantifying opinion fragmentation when in occurs.
To investigate these model behaviors, we compute the convergence time and the number of steady-state opinion clusters.
It is common to study these quantities in investigations of BCMs \cite{noorazar-et-al2020, meng2018, peralta2022}.

In some situations, an opinion cluster has very few agents. Consider a 500-node network in which 499 agents eventually have the same opinion, but the remaining agent 
(say, Agent 86, despite repeated attempts by Agent 99 and other agents to convince them) retains a distinct opinion at steady state.
In applications, it is not appropriate to think of this situation as opinion fragmentation. 
To handle such situations, we use a notion of ``major clusters'' and ``minor clusters'' \cite{laguna2004, lorenz2008}. We characterize major and minor clusters in an ad hoc way.
We define a ``minor'' opinion cluster in a network as an opinion cluster with at most 2\% of the agents. Any opinion cluster that is not a minor cluster is a ``major'' cluster. 
In our simulations, we calculate the numbers of major and minor opinion clusters at steady state. We only account for the number of major clusters when determining if a simulation reaches a consensus state (i.e., exactly one major cluster) or a fragmented state (i.e., more than one major cluster).
We still track the number of minor clusters and use the minor clusters when quantifying opinion fragmentation.

\begin{figure}[b]
\includegraphics[width=0.46\textwidth]{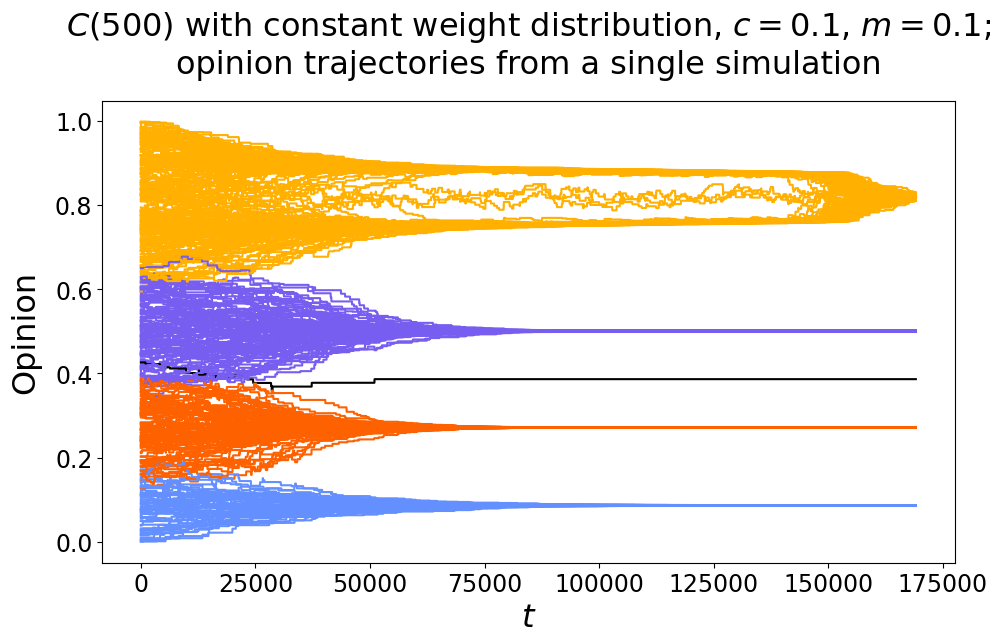}
\caption{Sample trajectories of agent opinions versus time in a single simulation of our node-weighted BCM on a 500-node complete graph with a constant weight distribution. Therefore, this situation corresponds to our baseline DW model. We color the trajectory of each node by its final opinion cluster.
Observe that the final opinion clusters have different sizes. There is a minor cluster (in black); it consists of a single node whose final opinion is about $0.4$. The opinion cluster that converges to the largest opinion value has about twice as many nodes as the other major clusters.
}
\label{fig:cluster_size} 
\end{figure}

Quantifying opinion fragmentation is much less straightforward than determining whether or not there is fragmentation. 
Researchers have proposed a variety of notions of fragmentation and polarization~\cite{bramson2016}, and they have also proposed several ways to quantify such notions~\cite{bramson2016, musco2021, adams2022}. In principle, a larger number of opinion clusters is one indication of more opinion fragmentation. However, as we show in Fig.~\ref{fig:cluster_size}, there can be considerable variation in the sizes (i.e., the number of nodes) of the opinion clusters. 
For example, suppose that there are two opinion clusters. If the two opinion clusters have the same size, then one can view the opinions in the system as more polarized than if one opinion cluster has a large majority of the nodes and the other opinion cluster has a small minority. 
Additionally, although we use only major clusters to determine if a system reaches a consensus or fragmented state, we seek to distinguish quantitatively between the scenarios of opinion clusters (major or minor) with similar sizes from ones with opinion clusters with a large range of sizes. 
Following Han et al.~\cite{han2020}, we do this by calculating Shannon entropy.

Suppose that there are $K$ opinion clusters, which we denote by $S_r$ for $r \in \{1, \ldots, K\}$.
We refer to the set $\{S_r\}_{r=1}^K$ as an ``opinion-cluster profile''; such a profile is a partition of a network.
The fraction of agents in opinion cluster $S_r$ is $|S_r|/N$. The Shannon entropy $H$ of the opinion-cluster profile is
\begin{equation}\label{eq:entropy}
    H = - \sum_{r = 1}^{K} \frac{|S_r|}{N} \ln \left( \frac{|S_r|}{N} \right) \,.
\end{equation}
The Shannon entropy $H$ gives us a scalar value to quantify the distribution of opinion-cluster sizes. 
For a given opinion-cluster profile, $H$ indicates the increase in information of knowing the opinion-cluster membership of a single agent instead of not knowing the cluster membership of any agents.
For a fixed $K$, the entropy $H$ is larger if the cluster sizes are closer in magnitude than if there is more heterogeneity in the cluster sizes. For opinion-cluster profiles with similar cluster sizes, $H$ is larger if there are more clusters. We use $H$ to quantify opinion fragmentation, with larger $H$ corresponding to more opinion fragmentation. 
We calculate the steady-state entropy $H(T)$ using all steady-state opinion clusters (i.e., both major and minor clusters).

Another way to quantify opinion fragmentation is to look at a local level and consider individual agents of a network. As Musco et al.~\cite{musco2021} pointed out, if an individual agent has many neighbors with with similar opinions to it, then it may be ``unaware'' of other opinions in the network. 
For example, an agent can observe that a majority of its neighbors hold an opinion that is uncommon in the global network. This phenomenon is sometimes called a ``majority illusion''~\cite{lerman2016}.
If a set of adjacent agents tend to have neighbors with similar opinions as theirs, they may be in an ``echo chamber'' \cite{flaxman2016}, as it seems that they are largely exposed only to conforming opinions.
To quantify the local observations of agents, Musco et al.~\cite{musco2021} calculated a notion of local agreement that measures the fraction of an agent's neighbors with opinions that are on the same side of the mean opinion in a network.
In our simulations, we often observe opinion fragmentation with three or more opinion clusters. 
Therefore, we need to look beyond the mean opinion of an entire network. To do this, we introduce the ``local receptiveness'' of an agent.
At time $t$, a node $i$ with neighborhood $\mathcal{N}(i)$ has a local receptiveness of
\begin{equation}\label{eq:local_receptiveness}
    L_i(t) = \frac{\left| \{j \in \mathcal{N}(i) : |x_i(t) - x_j(t)| < c  \} \right|}{|\mathcal{N}(i)|} \, .
\end{equation}
That is, $L_i(t)$ is the fraction of the neighbors of agent $i$ at time $t$ to which it is receptive (i.e., with which it will compromise its opinion if they interact).
In the present paper, we only consider connected networks, so each agent $i$ has $|\mathcal{N}(i)| \geq 1$ neighbors.
If one wants to consider isolated nodes, one can assign them a local receptiveness of $0$ or $1$.
In our numerical simulations, we calculate the local receptiveness of each agent of a network at the convergence time $T$. 
We then calculate the mean $\langle L_i(T) \rangle$ of all agents in the network. This is the steady-state mean local receptiveness, as it is based on edges in the steady-state effective-receptivity network $G_{\mathrm{eff}}(T)$. When consensus is not reached, a smaller mean local receptiveness is an indication of greater opinion fragmentation.
As we will discuss in Sec.~\ref{sec:results}, the Shannon entropy and the mean local receptiveness can provide insight into the extent of opinion fragmentation when one considers them in concert with the number of opinion clusters.


\section{Numerical simulations and results} \label{sec:results}

In this section, we present results of our numerical simulations of our node-weighted BCM. 
In our numerical experiments, the compromise parameter takes the values $m \in \{0.1, 0.3, 0.5\}$. For the confidence bound, we first consider the values $c \in \{0.1, 0.3, 0.5, 0.7, 0.9\}$, and we then examine additional values of $c$ near regions with interesting results.
As we discussed in Sec.~\ref{sec:sims}, for each individual graph, we simulate a total of 100 distinct sets of initial opinions and node weights in Monte Carlo simulations of our BCM.
For each of the random-graph models (i.e., ER and SBM graphs), we generate 5 graphs. 
For the 500-node complete graphs, we simulate the 10 weight distributions in Table~\ref{tab:distributions}. 
Because of computation time, we consider the 90-10 distributions only on 500-node complete graphs.
For the other networks in Table~\ref{tab:networks}, we consider 7 distributions in total: the constant weight distribution, the 80-10 distributions, and the 80-20 distributions.

In Table~\ref{tab:trends}, we summarize the trends that we observe in the examined networks. In the following subsections, we discuss details of our results for each type of network. 
The numbers of major and minor clusters, Shannon entropies, and values of mean local receptiveness are all steady-state values.
We include our code and figures in our repository at \url{https://gitlab.com/graceli1/NodeWeightDW}. In the present paper, we visualize our results using heat maps; in our code repository, we also show visualizations with line plots.

\newcommand{\trendleftcol}{0.5in}
\newcommand{\trendrightcol}{2.35in}
\newcommand{\toptrendvspace}{3pt} 
\newcommand{\bottomtrendvspace}{1pt} 
\newcommand{\trendvspace}{6pt}

\begin{table}
\caption{\label{tab:trends} Summary of the trends in our simulations of our node-weighted BCM. Unless we note otherwise, we observe these trends for each of the networks that we examine (complete graphs, ER and SBM random graphs, and the Caltech Facebook network).}
\begin{ruledtabular}
\begin{tabular}{m{\trendleftcol} m{\trendrightcol}}
Quantity & Trends \\ \hline
\begin{tabular}{m{\trendleftcol}} Convergence Time \end{tabular}
& \begin{tabular}{m{\trendrightcol}}
    \vspace{\toptrendvspace}
    $\bullet$ 
    For fixed values of $c$ and $m$,
    the heterogeneous weight distributions have longer convergence times than the constant weight distribution. 
    \vspace{\bottomtrendvspace}
    \end{tabular} \\ \hline
\begin{tabular}{m{\trendleftcol}}
Opinion \\ Fragmentation\footnotemark[1]
\end{tabular} &
\begin{tabular}{m{\trendrightcol}}
    \vspace{\toptrendvspace}
    $\bullet$ 
    For fixed values of $c \in [0.1, 0.4]$ and $m$,
    the heterogeneous weight distributions usually have more opinion fragmentation than the constant weight distribution. \\ \vspace{\trendvspace}
    $\bullet$ 
    For fixed values of $c$ and $m$ and
    a fixed distribution mean, 
    there is more opinion fragmentation as the tail of a distribution becomes heavier. \\ \vspace{\trendvspace}
    $\bullet$ 
    For fixed values of $c$ and $m$ and a given family of distributions, there is more opinion fragmentation when a distribution has a larger mean.
    \vspace{\bottomtrendvspace}
 \end{tabular}
\\ \hline
\begin{tabular}{m{\trendleftcol}}
Number of Major Clusters \end{tabular} 
& \begin{tabular}{m{\trendrightcol}}
    \vspace{\toptrendvspace}
    $\bullet$ 
    A larger minimum value of $c$ is required to always reach consensus for a heterogeneous weight distribution than for the constant weight distribution.
    \\ \vspace{\trendvspace}
    $\bullet$ 
    For fixed values of $c$ and $m$ and a fixed distribution mean, there are more major clusters as the tail of a distribution becomes heavier. \\ \vspace{\trendvspace}
    $\bullet$ 
    For fixed values of $c$ and $m$ and a given family of distributions, there are more major clusters when a distribution has a larger mean.
    \vspace{\bottomtrendvspace}
\end{tabular} \\ \hline
\begin{tabular}{m{\trendleftcol}}
Number of Minor Clusters \end{tabular}
& \begin{tabular}{m{\trendrightcol}}
    \vspace{\toptrendvspace}
    $\bullet$ For the constant weight distribution and for fixed $c$,
    there are typically more minor clusters when the compromise parameter $m \in \{0.3, 0.5\}$ than when $m = 0.1$.
    The heterogeneous weight distributions do not follow this trend.\footnotemark[2]  
    \vspace{\bottomtrendvspace}
    \end{tabular} \\
\end{tabular}
\end{ruledtabular}
\footnotetext[1]{We quantify opinion fragmentation using Shannon entropy and mean local receptiveness. We observe clearer trends for Shannon entropy than for the mean local receptiveness.}
\footnotetext[2]{For the Caltech network, we usually observe more minor clusters when $m \in \{0.3, 0.5\}$ than when $m = 0.1$ for each of our heterogeneous weight distributions.}
\end{table}


\subsection{Complete graphs} \label{sec:results-complete} 

The simplest underlying network structure on which we run our node-weighted BCM is a complete graph.
Complete graphs provide a baseline setting to examine how heterogeneous node-selection probabilities affect opinion dynamics.
In our numerical simulations on complete graphs, we consider all three means (which we denote by 80-10, 80-20, and 90-10 in Table~\ref{tab:distributions}) for each of the uniform, exponential, and Pareto node-weight distribution families.

The standard DW model on a complete graph with agents with opinions in the interval $[0,1]$ eventually reaches consensus if the confidence bound $c \geq 0.5$.
As one decreases $c$ from $0.5$, there are progressively more steady-state opinion clusters (both minor and major) \cite{lorenz2008, benaim2003}.
Lorenz \cite{lorenz2008} showed using numerical simulations that the number of major clusters is approximately
$\lfloor \frac{1}{2c} \rfloor$ for the standard DW model. 
Therefore, a transition between consensus and opinion fragmentation occurs for $c \in [0.25, 0.3]$. 
In our simulations, we observe that this transition occurs for $c \in [0.25, 0.4]$ in our node-weighted BCM.
To examine this transition, we thus zoom in on these values of $c$.
For the uniform and exponential distributions, we focus on $c \in [0.25, 0.3]$.
For the Pareto distributions, the transition occurs for larger values of $c$ than for the other distributions; we consider additional values of $c \in [0.3, 0.4]$.
Because the constant weight distribution is our baseline DW model, we simulate our BCM with the constant weight distribution for all values of $c$ that we consider for any other distribution.

\begin{figure}[ht!]
\includegraphics[width=0.48\textwidth]{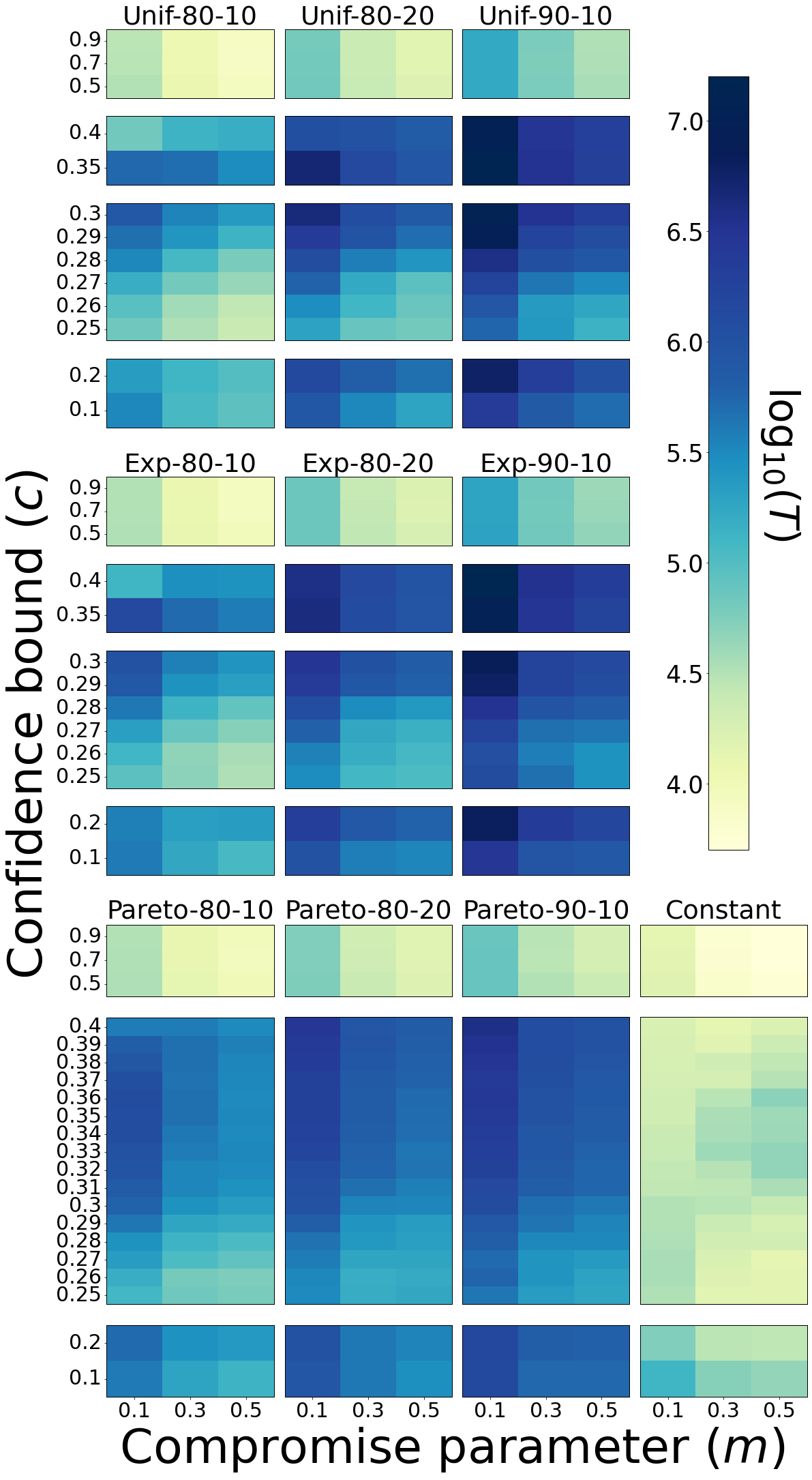}
\caption{Convergence times (in terms of the number of time steps) in simulations of our node-weighted BCM on a 500-node complete graph. If we only consider the time steps in which interacting nodes actually change their opinions, the convergence times are smaller; however, the trends are the same.
For this heat map and all subsequent heat maps, the depicted values are the means of our simulations of our BCM with each
node-weight distribution and each value of the BCM parameter pair $(c,m)$.
}
\label{fig:complete_T} 
\end{figure}

In Fig.~\ref{fig:complete_T}, we show the convergence times (which we measure in terms of the numbers of time steps) of our BCM simulations for various node-weight distributions. 
For fixed values of $c$ and $m$, all of the heterogeneous weight distributions yield longer convergence times than the constant weight distribution.
Additionally, for fixed $c$ and $m$ and a fixed family of distributions (uniform, exponential, or Pareto), the convergence time increases as we increase the mean of the distribution.
For fixed $c$ and for each heterogeneous weight distribution, the convergence time also increases as we decrease the compromise parameter $m$.
When calculating convergence time, we include time steps in which two nodes interact but do not change their opinions.
To see if the heterogeneous weight distributions have inflated convergence times as a result of having more of these futile interactions, we also calculate the number of time steps to converge when we exclude such time steps. That is, we count the total number of opinion changes that it takes to converge. On a logarithmic scale, there is little difference between the total number of opinion changes and the total number of time steps to converge.
We include a plot of the numbers of opinion changes in our
\href{https://gitlab.com/graceli1/NodeWeightDW}{code repository}.

\begin{figure}[ht!]
\includegraphics[width=0.48\textwidth]{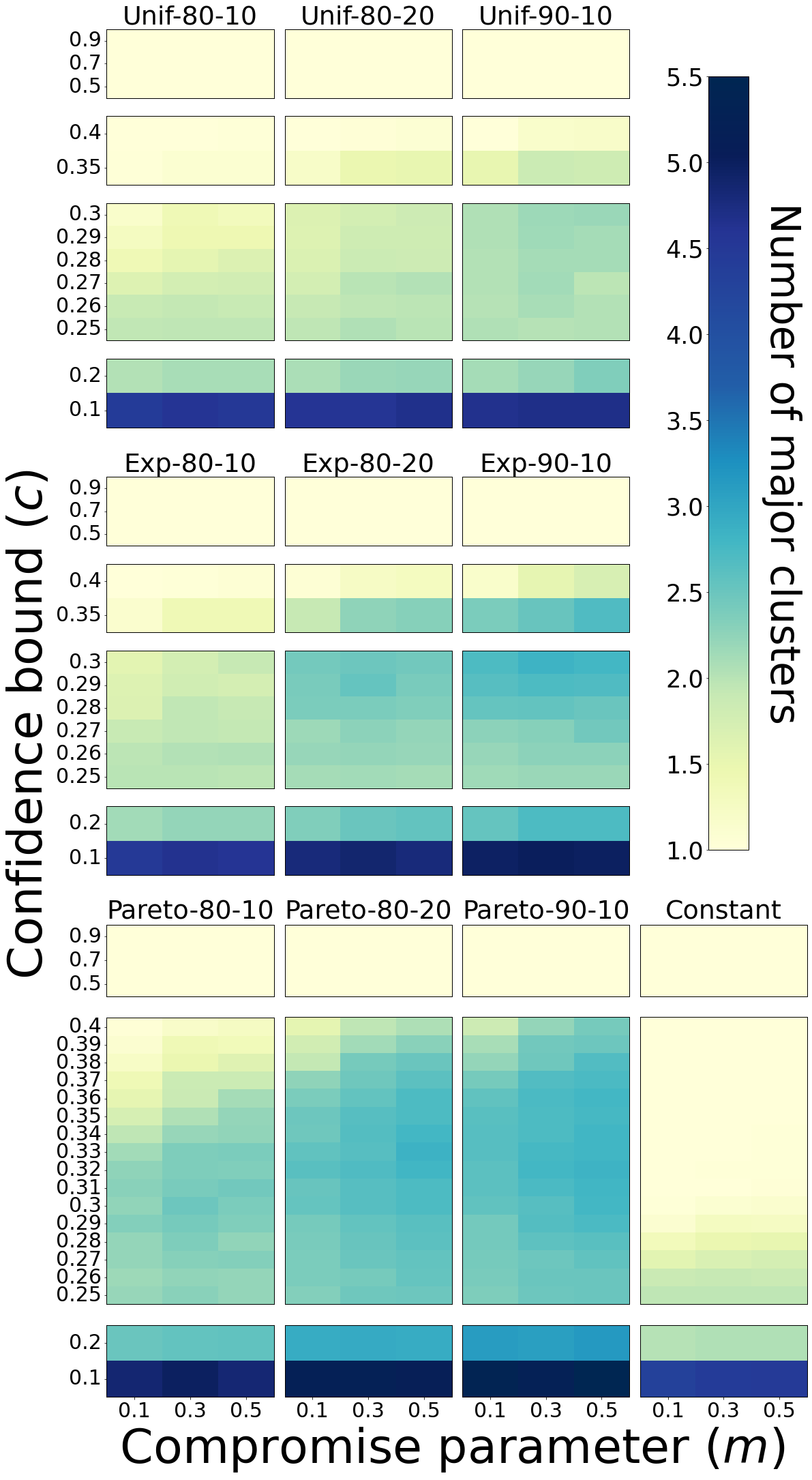}
\caption{\label{fig:complete_clusters} The numbers of major opinion clusters at steady state in simulations of our node-weighted BCM on a 500-node complete graph with various node-weight distributions. We consider a cluster to be major cluster if it has more than 2\% of the nodes of a network. (In this case, a major cluster must have at least 11 nodes.)}
\end{figure}

In Fig.~\ref{fig:complete_clusters}, we show the numbers of major opinion clusters at steady state in our BCM simulations for various node-weight distributions. For all weight distributions, consensus occurs in all of our simulations when the confidence bound $c \geq 0.5$.
For fixed values of $c \in [0.1, 0.4]$ and $m$, the heterogeneous weight distributions yield more steady-state major clusters than the constant weight distribution.
When we introduce heterogeneous node weights into our BCM, we need a larger confidence bound $c$ than for the constant weight distribution
to always reach consensus in our simulations.
It appears that our BCM with heterogeneous node weights
tends to have more opinion fragmentation than the baseline DW model.
For fixed $c$ and $m$, we observe for each distribution family (uniform, exponential, and Pareto) that there are more steady-state major clusters when the distribution mean is larger. 
To see this, proceed from left to right in Fig.~\ref{fig:complete_clusters} from the 80-10 distributions to the 80-20 distributions and then to the 90-10 distributions.
Additionally, for fixed values of $c$ and $m$ and a fixed distribution mean, there are more steady-state major clusters as we proceed from a uniform distribution to an exponential distribution and then to a Pareto distribution.

\begin{figure}[hb!]
\includegraphics[width=0.48\textwidth]{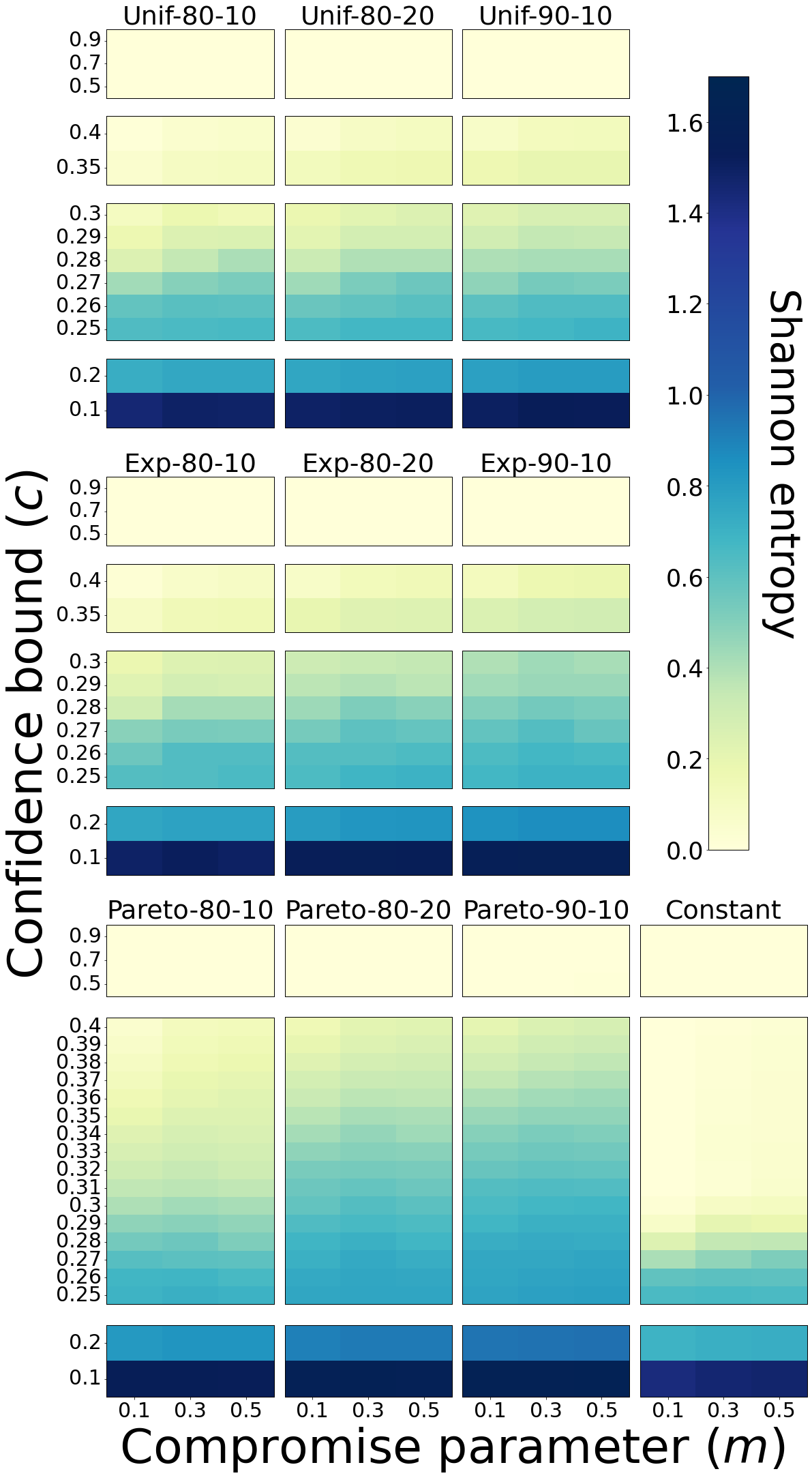}
\caption{\label{fig:complete_entropy} Shannon entropies of the steady-state opinion-cluster profiles in simulations of our node-weighted BCM on a 500-node complete graph with various node-weight distributions.}
\end{figure}

To investigate how the node-weight distribution and the BCM parameters (i.e., $c$ and $m$) affect the amount of opinion fragmentation, we calculate the Shannon entropy and mean local receptiveness (see Sec.~\ref{sec:quantities}) at steady state.
In Fig.~\ref{fig:complete_entropy}, we show the steady-state entropy values of our BCM simulations for various node-weight distributions.
For all node-weight distributions, when there is opinion fragmentation instead of consensus, the steady-state entropy increases as we decrease the confidence bound $c$ for fixed $m$.
In line with our observations in Fig.~\ref{fig:complete_clusters}, when $c \in [0.1, 0.4]$, simulations of heterogeneous weight distributions usually yield larger entropies than the constant weight distribution.
For fixed values of $c$ and $m$ and a fixed distribution mean,
we also observe a slightly larger entropy as we proceed from a uniform distribution to an exponential distribution and then to a Pareto distribution. 
For fixed $c$ and $m$, for the Pareto distributions, the entropy increases as we increase the mean of the distribution. (Proceed from left to right in Fig.~\ref{fig:complete_entropy}.)
The exponential and uniform distributions have the same trend, although it is less pronounced (i.e., the entropies do not increase as much) than for the Pareto distribution.
For the exponential and uniform distributions, a larger mean weight results in more major opinion clusters. 
For these two families of distributions, increasing the mean weight also tends to lead to smaller major opinion clusters.
Therefore, given either a uniform or an exponential distribution, we obtain similar Shannon entropies for different distribution means.
Consequently, if we quantify fragmentation using Shannon entropy, we conclude that in comparison to the Pareto distributions, increasing the mean weight has less effect on the amount of opinion fragmentation for the uniform and exponential distributions.
Because Shannon entropy depends on the sizes of the opinion clusters, it provides more information about opinion fragmentation than tracking only the number of major opinion clusters.
Our plot of the steady-state mean local receptiveness illustrates the same trends as the entropy. 
(See our \href{https://gitlab.com/graceli1/NodeWeightDW}{code repository} for the relevant figure.) This suggests that both Shannon entropy and mean local receptiveness are useful for quantifying opinion fragmentation.

We now discuss the numbers of steady-state minor opinion clusters in our BCM simulations on complete graphs. (See our \href{https://gitlab.com/graceli1/NodeWeightDW}{code repository} for a plot.)
For each node-weight distribution and each value of $c$ and $m$, when we take the mean of our 100 simulations, we obtain at most 2 steady-state minor clusters.
We observe the most minor clusters when $c \in \{0.1, 0.2\}$, which are the smallest confidence bounds that we examine.
For the constant weight distribution, we typically observe more minor clusters when $m \in \{0.3, 0.5\}$ than when $m = 0.1$.
However, we do not observe this trend for the heterogeneous weight distributions.
For example, for the Pareto-80-10 distribution, when $c \in [0.34, 0.4]$, decreasing $m$ results in more minor opinion clusters. 
For the Pareto distributions, as we decrease $m$, we also observe that minor clusters tend to appear at smaller confidence bounds. 
Smaller values of $m$ entail smaller opinion compromises for interacting agents; this may give more time for agents to interact before they settle into their final opinion clusters. 
For the constant weight distribution, this may reduce the number of minor clusters by giving more opportunities for agents to assimilate into a major cluster.
However, for our heterogeneous weight distributions, nodes with larger weights have a larger probability of interacting with other nodes and we no longer observe fewer minor clusters as we decrease $m$.

\begin{figure}[htb]
\includegraphics[width=0.48\textwidth]{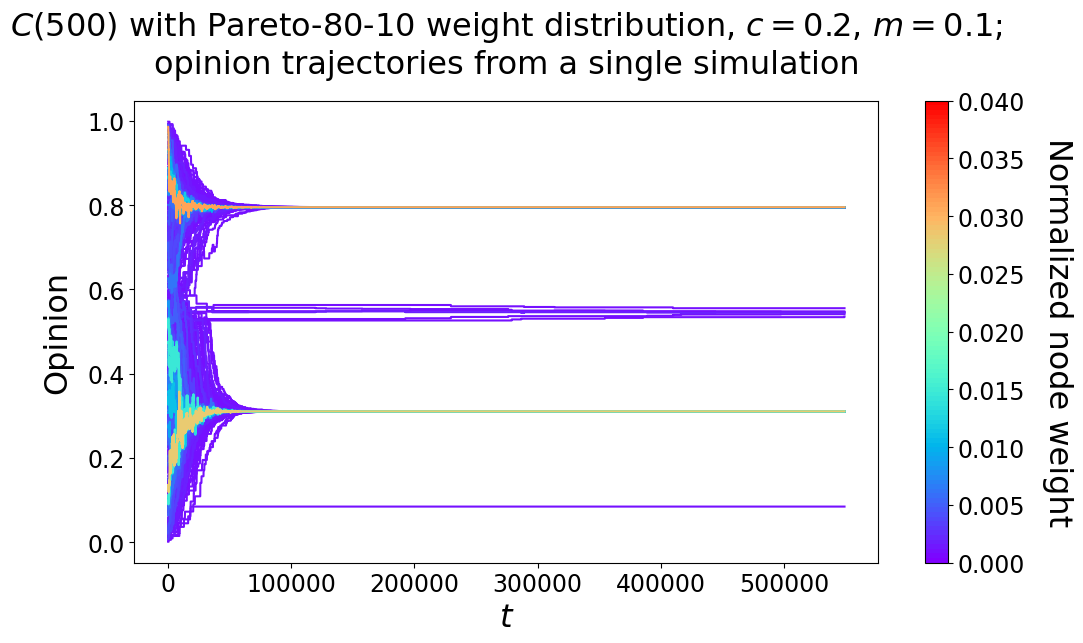}
\caption{Sample trajectories of agent opinions versus time in a single simulation of our node-weighted BCM on a complete graph with $N = 500$ nodes and node weights that we draw from a Pareto-80-10 distribution. We color the trajectory of each agent by its node weight, which we normalize so that the sum of all node weights is $1$. The nodes in the two minor opinion clusters are all small-weight nodes; their weights are close to $0$ (and are hence in purple).
}
\label{fig:pareto_trajectory}
\end{figure}

We now propose a possible mechanism by which our node-weighted BCM may promote the trends in Table~\ref{tab:trends}.
In Fig.~\ref{fig:pareto_trajectory}, we show the trajectories of opinions versus time for a single simulation with node weights that we draw from a Pareto-80-10 distribution.
To qualitatively describe our observations, we examine the large-weight and small-weight nodes (i.e., the nodes that are near and at the extremes of a set of node weights in a given simulation). Because our node-selection probabilities are proportional to node weights, to compare the weights in a simulation, we normalize them to sum to $1$. 
In Fig.~\ref{fig:pareto_trajectory}, the large-weight nodes appear to quickly stabilize into their respective steady-state major opinion clusters, and some small-weight nodes are left behind to form the two minor clusters.
In our numerical simulations on complete graphs, we observe that heterogeneity in the node weights results in large-weight nodes interacting more frequently than other nodes and quickly settling into steady-state major opinion clusters.
Small-weight nodes that are not selected for opinion updates early in a simulation are left behind to form the smallest clusters in a steady-state opinion-cluster profile; this increases the amount of opinion fragmentation.
In comparison to the constant weight distribution, when we increase the mean node weight or increase the relative proportion of large-weight nodes (by increasing the heaviness of the tail of the distribution) 
or decrease the value of the compromise parameter $m$, 
small-weight nodes take longer to settle into opinion clusters; this may promote both opinion fragmentation and the formation of minor opinion clusters.


\subsection{Erd\H{o}s--R\'{e}nyi (ER) graphs} \label{sec:results-ER} 

We now examine random graphs that we generate using $G(N,p)$ ER random-graph models, where $p$ is the homogeneous, independent probability of an edge between any pair of nodes \cite{newman2018}. For $p = 1$, these ER graphs are complete graphs.
In this subsection, we consider the edge probabilities $p \in \{0.1, 0.3, 0.5, 0.7\}$.

\begin{figure*}[th!]
\includegraphics[width=0.95\textwidth]{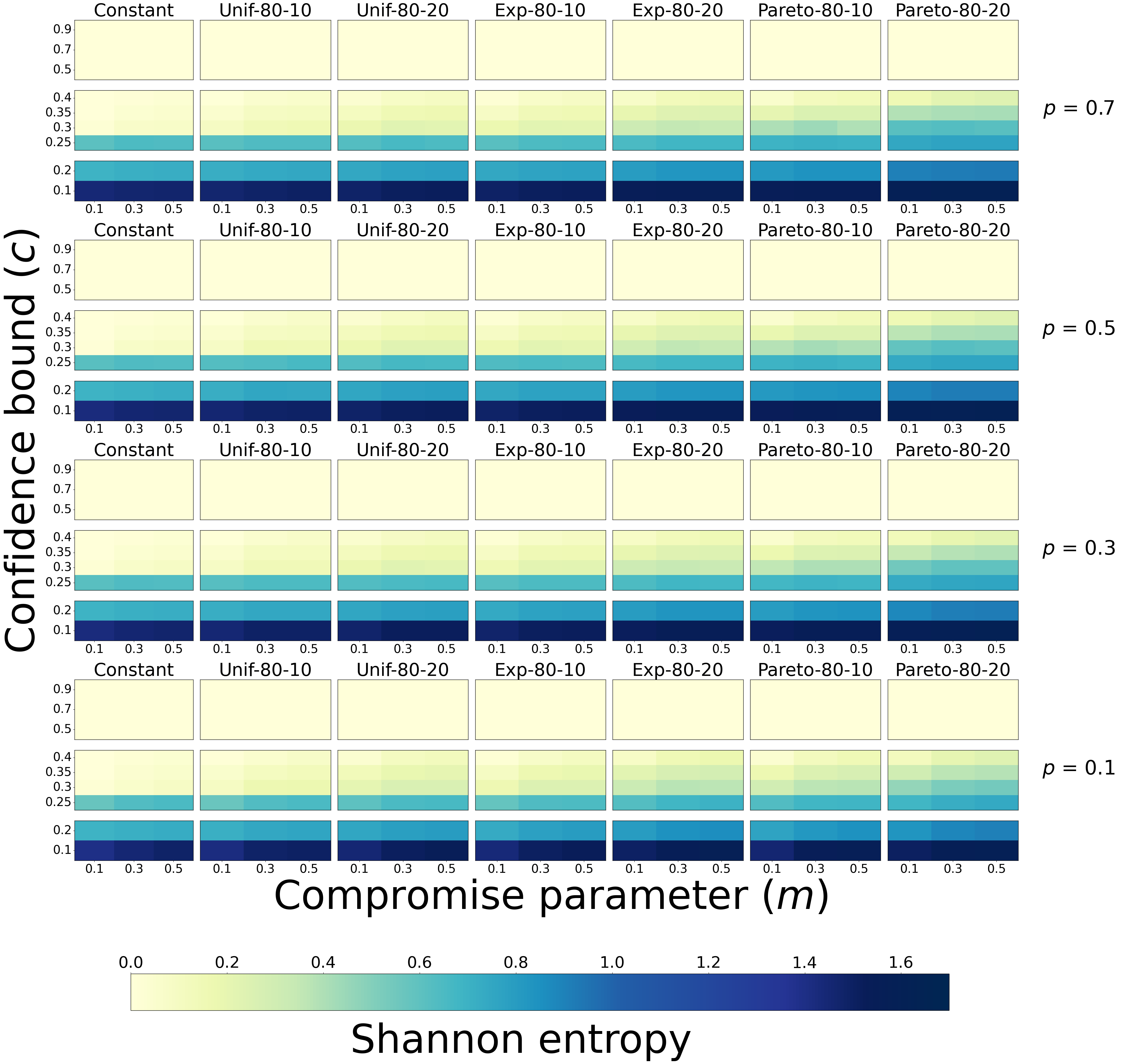}
\caption{\label{fig:er_entropy} Shannon entropies of the steady-state opinion-cluster profiles in simulations of our node-weighted BCM on $G(500, p)$ ER random graphs with various node-weight distributions.}
\end{figure*}

For each value of $p$, we observe the trends in Table~\ref{tab:trends}. We include the plots of our simulation results at steady state for the convergence times, the numbers of major and minor opinion clusters, and the values of mean local receptiveness in our \href{https://gitlab.com/graceli1/NodeWeightDW}{code repository}.
In Fig.~\ref{fig:er_entropy}, we show the steady-state Shannon entropies of our simulations for various node-weight distributions and values of $p$. 
The entropies are comparable to those that we obtained in our simulations on 500-node complete graphs (see Sec.~\ref{sec:results-complete}).
When $c \in [0.1, 0.4]$, for each of our three node-weight distribution families and for fixed values of $p$, $c$, and $m$, 
the 80-20 distribution tends to yield a larger Shannon entropy than the 80-10 distribution (which has a smaller mean).

For larger $p$, we expect the results of our simulations on $G(500, p)$ networks to be similar to those of our simulations on a 500-node complete graph.
For $p \in \{ 0.3, 0.5, 0.7\}$ and $N = 500$, the number of major opinion clusters and the mean local receptiveness are comparable to the corresponding results for a 500-node complete graph. 
When $p = 0.1$ and there is opinion fragmentation, for a fixed node-weight distribution and fixed values of $c$ and $m$, we observe fewer major opinion clusters than for larger values of $p$. 
For $p = 0.1$, when $c \in [0.1, 0.4]$, for a fixed node-weight distribution and fixed $c$ and $m$, we also observe that the mean local receptiveness tends to be larger than it is for larger $p$. 
One possible contributing factor for this observation may be that smaller values of $p$ yield $G(N,p)$ graphs with more small-degree nodes; these small-degree nodes have fewer available values of local receptiveness than larger-degree nodes.
For example, a node with degree $2$ can only have a local receptiveness of $0$, $0.5$, or $1$.
Unless a small-degree node is an isolated node in the steady-state effective-receptivity network $G_{\mathrm{eff}}(T)$, its presence may help inflate the value of the steady-state mean local receptiveness. 

For progressively smaller values of $p$, we observe progressively more minor opinion clusters at steady state. 
For $p \in \{0.5, 0.7\}$, the steady-state numbers of minor clusters are comparable to the numbers that we obtained for a 500-node complete graph.
When $p \in \{0.5, 0.7\}$, for each distribution and each value of $c$ and $m$, when we take the mean of our 500 simulations, we obtain at most 3 steady-state minor clusters. For these simulations, we observe the most minor clusters when $c \in \{0.1, 0.2\}$.
For $p = 0.1$, the mean number of minor clusters at steady state can be as large as $9$; this occurs when $c \in \{0.35, 0.4\}$.
It seems sensible that smaller values of $p$ yield more minor opinion clusters. For small $p$, there are more small-degree nodes than for larger values of $p$.
It is easier for small-degree nodes than for large-degree nodes to be in a minor opinion cluster, as small-degree nodes need to become unreceptive to few neighbors to end up in a minor cluster at steady state. That is, if $i$ is a small-degree node, few neighbors $j$ need to satisfy the inequality $|x_i - x_j| < c$.


\subsection{Stochastic-block-model (SBM) graphs} \label{sec:results-SBM} 

We now examine SBM random graphs that we generate using the parameters in Table~\ref{tab:networks}. For both the two-community and core--periphery SBM graphs, we observe the trends in Table~\ref{tab:trends}. 
We include the plots of our simulation results at steady state for the convergence times, the numbers of major and minor opinion clusters, the Shannon entropies, and the values of mean local receptiveness in our
\href{https://gitlab.com/graceli1/NodeWeightDW}{code repository}.

For the two-community SBM graphs, the steady-state Shannon entropies and numbers of major opinion clusters are comparable to those in our simulations on a complete graph. 
When there is opinion fragmentation, for a fixed node-weight distribution and fixed values of $c$ and $m$, the steady-state values of mean local receptiveness tend to be similar to the values for $G(500, 0.1)$ graphs and larger than the values for a complete graph.
The steady-state numbers of minor opinion clusters are similar to those for the $G(500, 0.1)$ random graphs. 

For the two-community SBM graphs, for each node-weight distribution and each value of $c$ and $m$, when we take the mean of our 500 simulations, we obtain at most 9 steady-state minor clusters. We observe the most steady-state minor clusters when $c \in \{0.35, 0.4\}$.
Recall that we select the edge probabilities of the two-community SBM so that each of the two communities has an expected mean degree that matches that of $G(500, 0.1)$ graphs.
Therefore, it is reasonable that we obtain similar results for the two-community SBM and the $G(500, 0.1)$ random graphs. 
In our numerical simulations, we assign the node weights randomly without considering the positions (which, in this case, is the community assignments) of the nodes of a network. 
With node weights assigned in this way,
it seems that graph sparsity may be more important than community structure for determining if the system reaches a consensus or fragmented state.

For a fixed node-weight distribution and fixed values of $c$ and $m$, the core--periphery SBM graphs tend to have fewer major clusters than complete graphs. 
Additionally, both the steady-state Shannon entropy and the mean local receptiveness tend to be larger for the core--periphery SBM graphs than for complete graphs.
Larger entropy and smaller local receptiveness are both indications of more opinion fragmentation. 
If we consider only the number of major opinion clusters, it seems that the core--periphery SBM graphs yield less opinion fragmentation than complete graphs. 
However, when we examine the entire opinion-cluster profile of a network and account for the cluster sizes and the minor clusters, the Shannon entropy reveals that there is more opinion fragmentation in core--periphery SBM graphs than in complete graphs. 
The steady-state mean local receptiveness indicates that the nodes of a core--periphery SBM graph tend to be receptive to a larger fraction of their neighbors than the nodes of a complete graph.

We believe that Shannon entropy provides a more useful quantification than mean local receptiveness of opinion fragmentation in a network. For networks with a large range of degrees, small-degree nodes can inflate the mean value of local receptiveness.
Analogously, for clustering coefficients, a network's mean local clustering coefficient places more importance on small-degree nodes than its global clustering coefficient \cite{newman2018}. 
In the context of our node-weighted BCM, consider a node with degree 2 and a node with degree 100, and suppose that both of them have a local receptiveness of 0.5. The larger-degree node having a local receptiveness of 0.5 gives a better indication that there may be opinion fragmentation in a network than the smaller-degree node having the same local receptiveness. 
However, we treat both nodes equally when we calculate the mean local receptiveness. 
We believe that local receptiveness is a useful quantity to calculate for individual nodes to determine how they perceive the opinions of their neighbors. However, it appears to be less useful than Shannon entropy for quantifying opinion fragmentation in a network.

For a fixed node-weight distribution and fixed values of $c$ and $m$,
the steady-state numbers of major opinion clusters that we obtain in the core--periphery SBM graphs are comparable to the numbers for a complete graph.
The steady-state numbers of minor opinion clusters tend to be larger for core--periphery SBM graphs than for two-community SBM graphs (which have more minor clusters than a complete graph). 
For each node-weight distribution and each value of $c$ and $m$, when we take the mean of our 500 simulations, we observe at most 11 steady-state minor clusters; this occurs when $c = 0.1$. 
One possibility is that the core--periphery structure makes it easier to disconnect peripheral nodes of an effective-receptivity network, causing these nodes to form minor clusters. 
For the core--periphery SBM graphs, it seems interesting to investigate the effect of using network structure to assign which nodes have large weights.
For example, if we assign all of the large weights to nodes in the core, will that pull more of the peripheral nodes into opinion clusters with core nodes? If we place a large-weight node in the periphery, will it be able to pull core nodes into its opinion cluster?


\subsection{Caltech network} \label{sec:results-Caltech}

\begin{figure*}[htb!]
\includegraphics[width=0.93\textwidth]{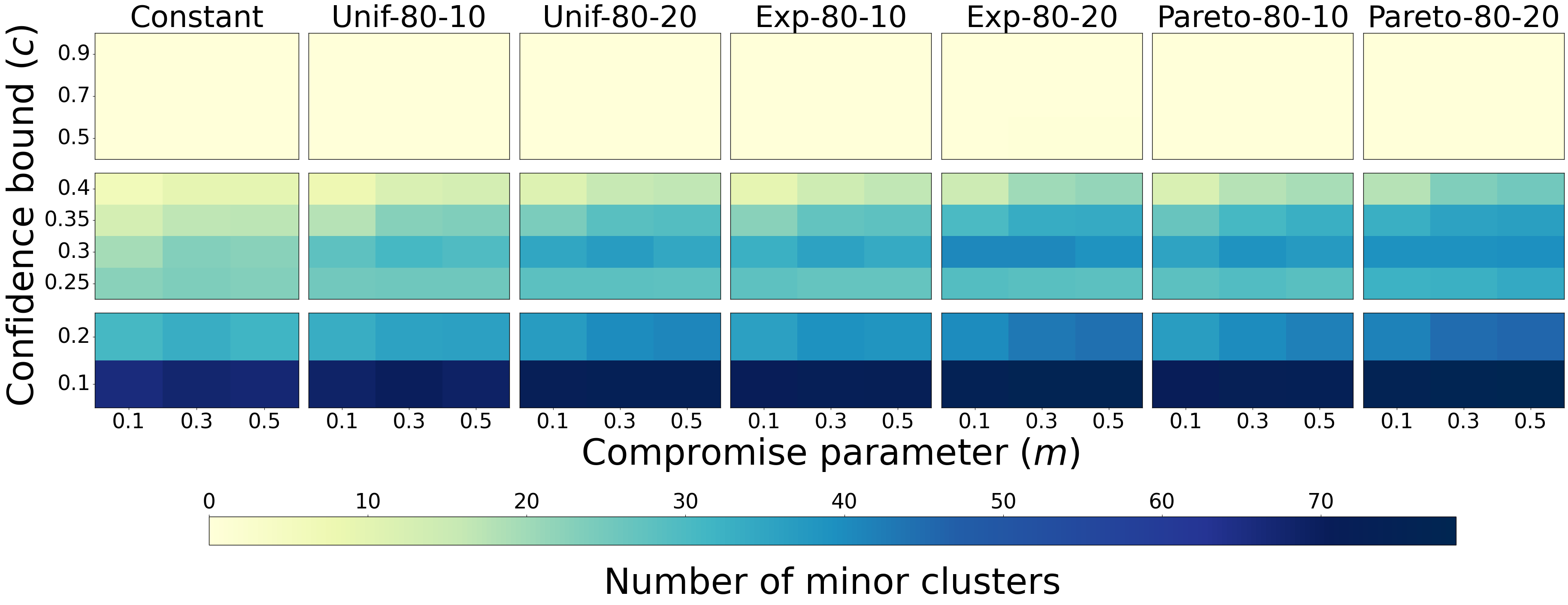}
\caption{\label{fig:caltech_minor} The steady-state numbers of minor opinion clusters in simulations of our node-weighted BCM on the Caltech Facebook network with various distributions of node weights. We consider a cluster to be minor cluster if it has at most 2\% of the nodes (i.e., 15 or fewer nodes) of a network.}
\end{figure*}

\begin{figure*}[htb!]
\includegraphics[width=0.93\textwidth]{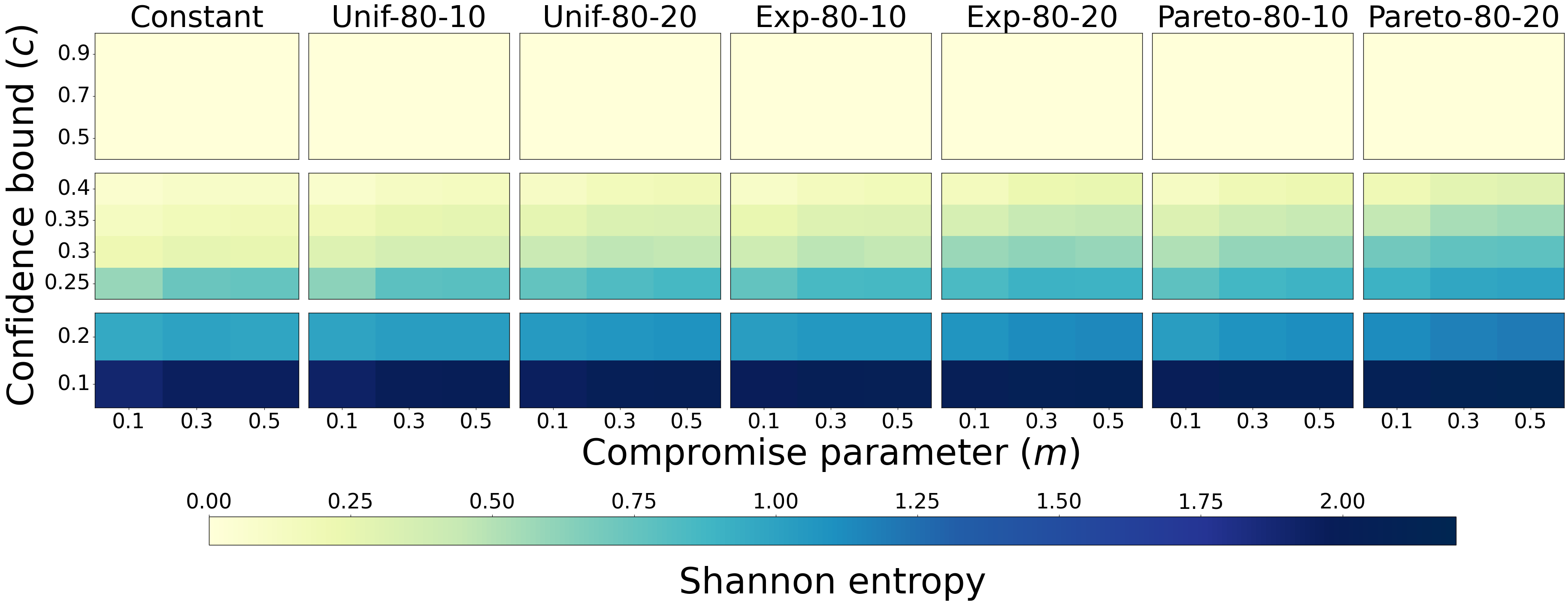}
\caption{\label{fig:caltech_entropy} Shannon entropies of the steady-state opinion-cluster profile in simulations of our node-weighted BCM on the Caltech Facebook network with various node-weight distributions.}
\end{figure*}

We now discuss the Caltech Facebook network, which is an empirical data set in which the nodes are individuals with Caltech affiliations and the edges represent ``friendships'' on Facebook on one day in fall 2005 \cite{red2011, traud2012}. We consider the network's largest connected component, which has 762 nodes and 16,651 edges. 
The Caltech network has all but one of the trends that we reported in Table~\ref{tab:trends}; the only exception is the trend in the number of minor opinion clusters.
When there is opinion fragmentation, the Caltech network has more steady-state minor clusters and larger steady-state Shannon entropies than in the synthetic networks.

In Fig.~\ref{fig:caltech_minor}, we show the steady-state numbers of minor opinion clusters in simulations of our BCM on the Caltech network. We obtain the most minor clusters when $c = 0.1$, which is the smallest value of $c$ that we examine. 
For each node-weight distribution and each value of $c$ and $m$, when we take the mean of our 100 simulations on the Caltech network, we obtain as many as 78 minor clusters, which is much more than the single-digit numbers that we usually observe for our synthetic networks.
Additionally, unlike in our synthetic networks, for all distributions (not just the constant weight distribution), the Caltech network tends to have more minor clusters when $m \in \{0.3, 0.5\}$ than when $m = 0.1$. 
We include our plot of the steady-state number of major opinion clusters in our \href{https://gitlab.com/graceli1/NodeWeightDW}{code repository}.
The Caltech network tends to have fewer major opinion clusters than the examined synthetic networks.

In Fig.~\ref{fig:caltech_entropy}, we show the steady-state Shannon entropies for the Caltech network.
For a fixed node-weight distribution and fixed values of $c$ and $m$,
when there is opinion fragmentation, the Caltech network has a larger entropy than for our synthetic networks.
This aligns with our observation that the Caltech network has many more minor opinion clusters than our synthetic networks. 
We show a plot of the steady-state values of mean local receptiveness for the Caltech network in our \href{https://gitlab.com/graceli1/NodeWeightDW}{code repository}.
The values of the mean local receptiveness tend to be larger for the Caltech network than for the 500-node complete graph. We suspect that this arises from the presence of many small-degree nodes in the Caltech network. In Sec.~\ref{sec:results-SBM}, we discussed the impact of small-degree nodes on the mean local receptiveness.

\begin{figure}[htb!]
\includegraphics[width=0.35\textwidth]{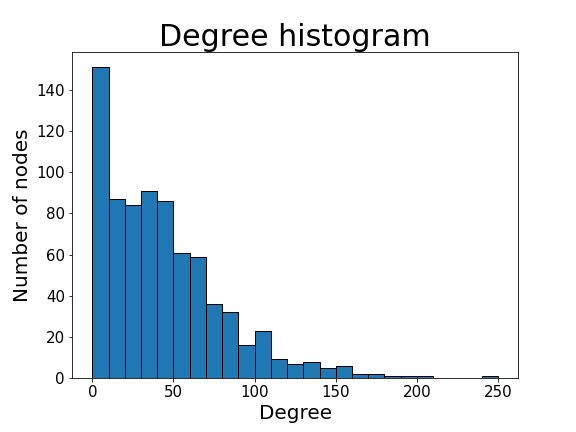}
\caption{\label{fig:caltech_degree} Histogram of the node degrees of the Caltech Facebook network. The bins have width 10 and originate at the left end point (i.e., the bins indicate degrees of 0--9, 10--19, and so on).}
\end{figure}

The histogram of the node degrees of the Caltech network (see Fig.~\ref{fig:caltech_degree}) differs dramatically from those of our synthetic networks.
Unlike in our synthetic networks, the most common degrees in the Caltech network are among the smallest degrees. In Fig.~\ref{fig:caltech_degree}, the tallest bar in the histogram is for nodes of degrees 0--9. These abundant small-degree nodes are likely to disconnect from the largest connected component(s) of the effective-receptivity network and form minor opinion clusters. 
Because we select the initial opinions uniformly at random from $[0,1]$, when $c = 0.1$, it is possible that small-degree nodes are initially isolated nodes of the effective-receptivity network because of their initial opinions. The abundance of small-degree nodes in the Caltech network helps explain its larger steady-state numbers of minor opinion clusters and the correspondingly larger 
entropies than for our synthetic networks.
Despite the fact that the Caltech network is structurally very different from our synthetic networks, it follows all of the trends in Table~\ref{tab:trends} aside from the one for the number of minor opinion clusters. 
Therefore, it seems that the trends that we observe in our node-weighted BCM when we assign node weights uniformly at random (and hence in a way that is independent of network structure) are fairly robust to the underlying network structure.


\subsection{Finite-size effects} \label{sec:results-finite_size} 
We now investigate finite-size effects in our BCM results for our simulations on a complete graph. 
To ensure reasonable computation times, we examined synthetic networks with 500 nodes.
However, it is useful to get a sense of whether or not the trends in Table~\ref{tab:trends} hold for networks of different sizes. 
To start to investigate this, we simulate our BCM on complete graphs of sizes $N \in \{10, 20, 30, 45, 65, 100, 150, 200, 300, \ldots, 1000\}$.
We examine $m \in \{0.3, 0.5\}$, and $c \in \{0.1, 0.3, 0.5\}$, which give regimes of opinion fragmentation, a transition between fragmentation and consensus for the constant weight distribution, and opinion consensus.
We consider the constant weight distribution and the 80-10 distributions (i.e., the uniform, exponential, and Pareto distributions with a mean node weight of 2.8836). We do not examine any larger-mean distributions because they require longer computation times.

\begin{figure*}[htb!]
\includegraphics[width=0.95\textwidth]{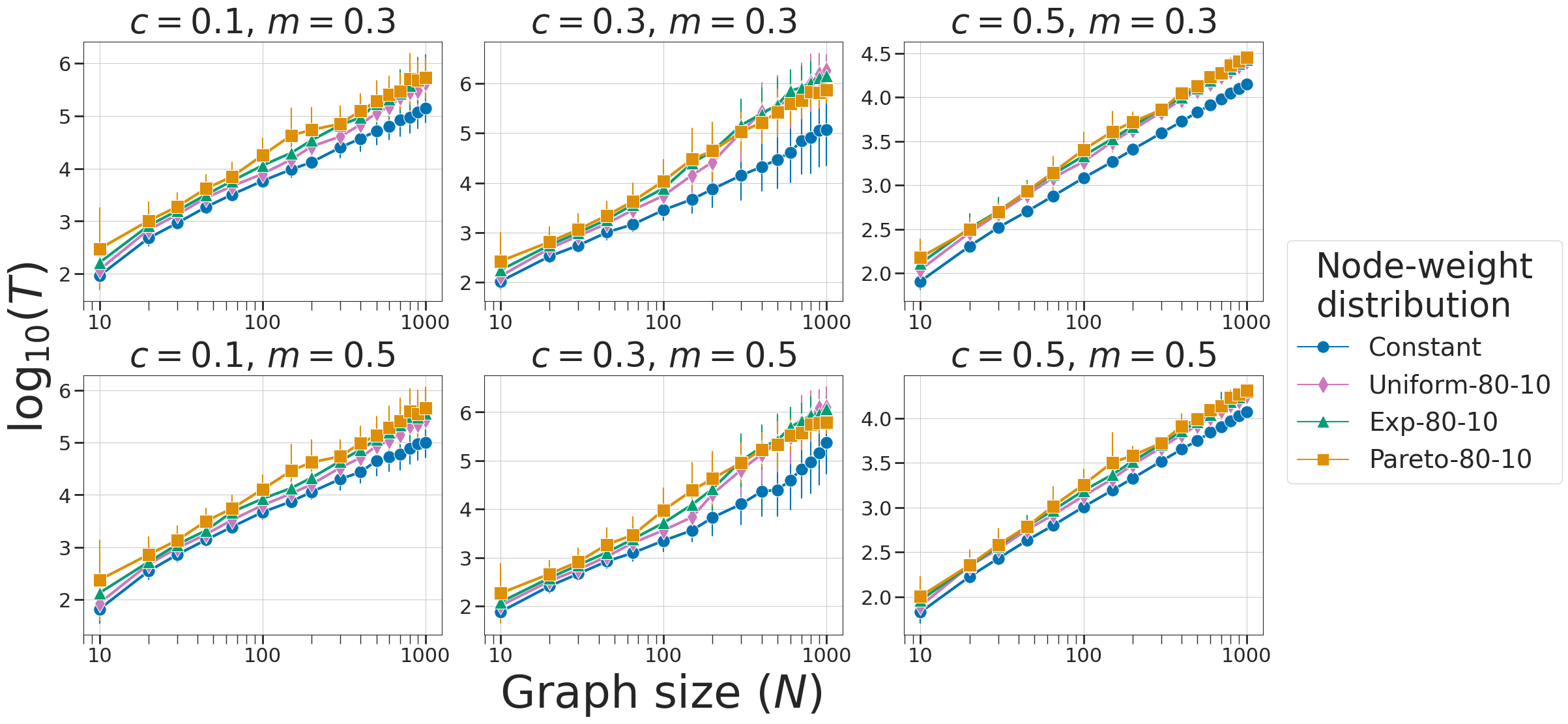}
\caption{Convergence times (in terms of the number of time steps) in simulations of our node-weighted BCM on complete graphs of various sizes. We show results for various choices of $c$ and $m$; the marker shape and color indicate the node-weight distribution. For this figure and subsequent figures of this type, the points are means of 100 simulations and the error bars indicate one standard deviation from the mean. 
The horizontal axis gives the graph size on a logarithmic scale. For clarity, the vertical axes of the plots have different scales.}
\label{fig:size_T} 
\end{figure*}

In Fig.~\ref{fig:size_T}, we show the convergence times of our simulations of our BCM on complete graphs of various sizes. 
To visualize our results, we plot the graph sizes on a logarithmic scale.
For all distributions, the convergence times become longer as we increase the graph size.
For each graph size, the convergence times for the heterogeneous weight distributions are similar to each other and are longer than those for the constant weight distribution.

\begin{figure*}[htb!]
\includegraphics[width=0.95\textwidth]{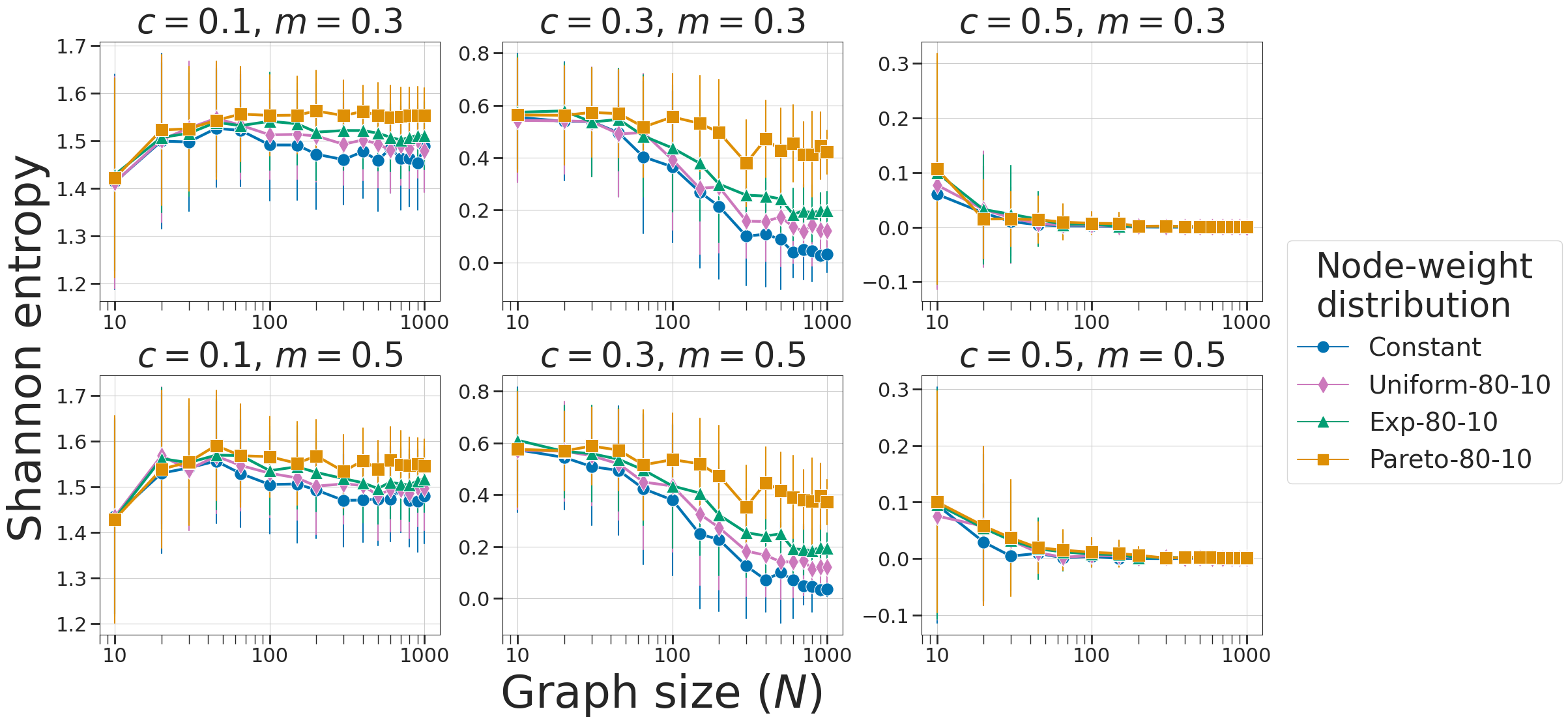}
\caption{Shannon entropies of the steady-state opinion-cluster profiles in simulations of our node-weighted BCM on complete graphs of various sizes. We show results for various choices of $c$ and $m$; the marker shape and color indicate the node-weight distribution.
}
\label{fig:size_entropy} 
\end{figure*}

In Fig.~\ref{fig:size_entropy}, we show the steady-state Shannon entropies from our simulations of our BCM on complete graphs of various sizes. For a fixed value of $c$, we observe similar results when $m = 0.3$ and $m = 0.5$. 
When $c = 0.5$, for each distribution, the simulations always reach a consensus (i.e., there is exactly one major steady-state opinion cluster) for $N \geq 200$. Correspondingly, the steady-state entropies are close to $0$. (They are not exactly $0$ because the calculation of Shannon entropies includes information from minor clusters.)
As we increase the network size, the error bars (which indicate one standard deviation from the mean) become progressively smaller.
When $c \in \{0.1, 0.3\}$, for sufficiently large graph sizes (specifically, when $N \geq 100$), we observe that the entropy increases as we increase the heaviness of the tail of a distribution.
For $c = 0.3$, the mean steady-state entropies appear to no longer change meaningfully with $N$ when $N \geq 400$. 
For $c = 0.1$, this is the case when $N \geq 100$.

When there is opinion fragmentation, the heterogeneous node-weight distributions yield larger steady-state Shannon entropies (and hence more opinion fragmentation, if one is measuring it using entropy) than the constant weight distribution for each graph size. 
Additionally, for a given distribution mean, we obtain larger entropies (and thus more opinion fragmentation) as we increase the heaviness of the tail of a distribution.
We have not explored the effect of graph size on the trends that we observe (see Table~\ref{tab:trends}) when we increase
the distribution mean for a fixed family of distributions.
In our \href{https://gitlab.com/graceli1/NodeWeightDW}{code repository},
we include a plot of the the steady-state mean local receptiveness for complete graphs of various sizes. In that plot, we also observe the trend of more opinion fragmentation (in the sense of a smaller mean local receptiveness) for heterogeneous node-weight distributions with increasingly heavy tails.

We also examine the steady-state numbers of major and minor opinion clusters in simulations of our BCM on complete graphs of various sizes; we include plots of them in our \href{https://gitlab.com/graceli1/NodeWeightDW}{code repository}.
For a fixed value of $c$, we observe similar results when $m = 0.3$ and $m = 0.5$. 
When $N \leq 49$, there are no minor opinion clusters, by definition, because minor clusters can include at most 2\% of the nodes of a network (and even a single node constitutes more than 2\% of all nodes for such small networks).
When $N \geq 65$ and $c \in \{0.1, 0.3\}$, for each distribution,
the number of minor clusters tends to increase as we increase $N$.
We do not observe a clear trend in which distributions yield more minor clusters. When $c = 0.5$, the mean number of minor clusters is always near $0$. 
When $c = 0.5$ and $N \geq 200$, all simulations yield $1$ major opinion cluster (i.e., they all reach consensus).
When $c = 0.3$, for all graph sizes, there are more major opinion clusters as we increase the heaviness of the tail of a distribution. Additionally, when $c = 0.3$, for the Pareto-80-10 distribution, the number of major clusters tends to increase as we increase the graph size. 
For the other distributions, the number of major clusters tends to decrease as we increase the graph size.
When $c = 0.1$ and $N \geq 200$, there again tends to be more major clusters as we increase the heaviness of the tail of a distribution, although the trend is not as clear as it was for $c = 0.3$.

For graphs with $N = 500$ or more nodes, the mean steady-state Shannon entropies for each node-weight distribution appear to no longer change meaningfully with respect to $N$; the mean entropies are more consistent for $N \geq 500$ than for smaller values of $N$.
For each graph size, the heterogeneous 80-10 distributions have longer convergence times than the constant weight distribution.
In all of these cases, we also observe more opinion fragmentation as we increase the heaviness of the tail of a distribution.
Because of computation time, we have not examined finite-size effects for different values of the distribution means.
However, because the mean Shannon entropies no longer change meaningfully with respect to $N$ for graphs with $N \geq 500$ nodes, we hypothesize that the trends in opinion fragmentation and convergence time in Table~\ref{tab:trends} continue to hold for our synthetic networks when there are more than 500 nodes.


\section{Conclusions and discussion} \label{sec:discussion} 

We developed a novel bounded-confidence model (BCM) with heterogeneous node-selection probabilities, which we modeled by using node weights. One can interpret these node weights as encoding phenomena such as heterogeneous agent sociabilities or activity levels.
We studied our node-weighted BCM with fixed node weights that we assign in a way that disregards network structure and node opinions.
We demonstrated that our node-weighted BCM has longer convergence times and more opinion fragmentation than a baseline Deffuant--Weisbuch (DW) BCM in which we uniformly randomly select nodes for interaction. It is straightforward to adapt our BCM to assign node weights in a way that depends on network structure and/or node opinions. See Sec.~\ref{sec:discussion-networks} and Sec.~\ref{sec:discussion-opinion} for discussions.


\subsection{Summary of our main results} \label{sec:discussion-summary} 

We simulated our node-weighted BCM with a variety of node-weight distributions (see Table~\ref{tab:distributions}) on several random and deterministic networks (see Table~\ref{tab:networks}). 
For each of these distributions and networks, we systematically investigated the convergence time and opinion fragmentation for different values of the confidence bound $c$ and the compromise parameter $m$. 
To determine if the nodes of a network reach consensus or if there is opinion fragmentation, we calculated the steady-state number of major clusters in our simulations. To quantify the amount of opinion fragmentation, we calculated the steady-state Shannon entropy and mean local receptiveness. 
For a given network, we found that entropy and mean local receptiveness follow the same trends in which distributions have more opinion fragmentation (see Table~\ref{tab:trends}).
Based on our results, we believe that Shannon entropy is more useful than mean local receptiveness for quantifying opinion fragmentation in a network.
However, calculating local receptiveness is insightful for explorations of the opinion dynamics of individual nodes.

In our simulations of our node-weighted BCM, we observed a variety of typical trends (see Table~\ref{tab:trends}). 
In particular, we found that heterogeneous node-weight distributions yield longer convergence times and more opinion fragmentation than the baseline DW model (which we obtain by using a constant weight distribution) in simulations of our BCM.
Opinion fragmentation also increases if either (1) for a fixed distribution mean, we make the tail of the distribution heavier or (2) for a given distribution family, we increase the mean of the distribution.
Given a set of heterogeneous node weights, we hypothesize that large-weight nodes are selected early in a simulation with large probabilities and quickly settle into their associated steady-state major opinion clusters.
Small-weight nodes that are not selected early in a simulation are left behind to form small opinion clusters, resulting in more opinion fragmentation than in the baseline DW model.


\subsection{Relating node weights to network structure} \label{sec:discussion-networks} 

We examined deterministic and random graphs with various structures, and we observed the trends in Table~\ref{tab:trends}.
For each of our BCM simulations, we selected node weights from a specified distribution and then assigned these weights to nodes uniformly at random.
Therefore, our investigation conveys what trends to expect with fixed, heterogeneous node weights that are assigned to nodes without regard for network structure. 
However, our model provides a flexible framework to study the effects of node weights when they are correlated with network structure. 
For example, one can assign weights to nodes in a way that depends on some centrality measure (such as degree).
In our BCM, we expect large-degree and large-weight nodes to have more interactions than small-degree or small-weight nodes. Nodes with larger degrees have more neighbors that can select them for an interaction, and nodes with larger weights have associated larger probabilities of being selected for an interaction.
One possible area of future work is to investigate the combined effects of node weight and node degree on the frequency of interactions and the distribution of steady-state opinions in our BCM.
Mean-field approaches, such as the one in \cite{fennell2021}, may offer insights into these effects.

For a given set of node weights, larger-weight nodes have larger probabilities of interacting with other nodes; their position in a network likely affects the dynamics of BCMs and other models of opinion dynamics.
One can also investigate the effects of homophily in the assignment of node weights. 
For example, in social-media platforms, very active accounts may engage with each other more frequently by sharing or commenting on each others' posts. 
We can incorporate such features into our BCM through a positive node-weight assortativity, such that large-weight nodes are more likely to be adjacent to each other than to other nodes.

As in the standard DW model, we assign the initial opinions uniformly at random in our BCM. However, in a real social network with community structure, this choice may not be realistic. 
One can investigate a social network with communities with different mean opinion values and examine the effect of placing large-weight nodes into different communities. For example, how does placing all large-weight nodes into the same community affect opinion dynamics and steady-state opinion-cluster profiles?
How does the presence of a small community of ``outspoken'' (i.e., large-weight) nodes influence the final opinions of nodes in other communities of a network?
Will the small community quickly engender an echo chamber \cite{flaxman2016}, will it pull other nodes into its final opinion cluster, or will something else occur? 


\subsection{Relating node weights to node opinions} \label{sec:discussion-opinion} 

In the present paper, we considered fixed node weights that are independent of node opinions. One can readily adapt our BCM to incorporate time-dependent node weights, such as ones that depend on node opinions. One can allow the probability of selecting a node for interaction to depend on how extreme its opinion is \cite{alizadeh2015} or on the similarity of its opinion to that of another node \cite{sirbu2019}.

S\^{i}rbu et al.~\cite{sirbu2019} studied a modified DW model with heterogeneous node-selection probabilities that model algorithmic bias on social media.
In their model, one first selects an agent uniformly at random. One then calculates the magnitude of the opinion difference between that agent and each of its neighbors and then selects a neighbor with a probability that is proportional to this difference. 
In the context of our BCM, one can represent their agent-selection mechanism using time-dependent node weights.
To do this, at each time $t$, one first assigns the same constant weight to all nodes when selecting a first node $i$.
When selecting a second node $j$ to interact with $i$, one then assign weights to neighbors of $i$ that are a function of the opinion difference $|x_i(t) - x_j(t)|$.
One assigns a weight of $0$ to nodes that are not adjacent to $i$.
The simulations by S\^{i}rbu et al. on complete graphs suggest that greater algorithmic bias results in longer convergence times and more opinion clusters \cite{sirbu2019}.
Very recently, Pansanella et al.~\cite{pansanella2022} observed similar trends in a study of the algorithmic-bias model of S\^{i}rbu et al. for various random-graph models.

In our simulations of our BCM with heterogeneous node-selection probabilities, we observed similar trends of longer convergence times and more opinion clusters (and opinion fragmentation) than in our baseline DW model. Our results illustrate that it is important to consider the baseline effect of assigning node weights uniformly at random in studies of BCMs with heterogenous node-selection probabilities before attributing trends such as longer convergence times and more opinion fragmentation to specific mechanisms such as algorithmic bias. Different mechanisms can yield very similar empirical observations.


\subsection{Edge-based heterogeneous activities} \label{sec:discussion-edges} 

In the standard DW model, at each time, one selects an edge of a network uniformly at random and the two agents that are attached to that edge interact with each other \cite{weisbuch2001}.
Most past work on the DW model and its extensions has focused on this edge-based selection mechanism \cite{noorazar2020}.
In our BCM, to incorporate node weights (e.g., to encode heterogeneous sociabilities or activity levels of individuals), we instead used a node-based selection mechanism.
For voter models of opinion dynamics, it is known that the choice between edge-based and node-based agent selection can substantially affect a model's qualitative behavior \cite{kureh2020}. 
We are not aware of a comparison of edge-based and node-based agent selection in asynchronous BCMs (and, in particular, in DW models), and it seems interesting to investigate this issue. 

We developed our BCM to incorporate node weights that encode 
heterogeneous activity levels of individuals.
One can also examine heterogeneous dyad-activity levels to account for the fact that individuals do not interact with each of their social contacts with the same probability.
To encode such heterogeneity, one can construct a variant of our BCM that incorporates edge weights.
At each time step, one can select a pair of agents to interact with a probability that is proportional to weight of the edge between them. 
We have not yet examined edge-based heterogeneous activity levels in a BCM, and we expect that it will be interesting to investigate them.


\subsection{Importance of node weights} \label{sec:discussion-weights} 

The key novelty of our BCM is our incorporation of node weights into opinion dynamics. 
Node weights have been used in activity-driven models of temporal networks \cite{perra2012}, and activity-driven frameworks have been used to model 
which agents can interact with each other in models of opinion dynamics \cite{li2017,zhang2018}.
In our BCM, the node weights determine the probabilities to select agents for interaction in a time-independent network.
Alizadeh and Cioffi-Revilla \cite{alizadeh2015}, S\^{i}rbu et al. \cite{sirbu2019}, and Pansanella et al. \cite{pansanella2022} examined specific scenarios of heterogeneous node-selection probabilities in DW models. 
Our node-weighted BCM provides a general framework to incorporate node weights into an asynchronous BCM. Using our framework, one can consider node weights that fixed and are assigned uniformly at random to nodes (i.e., as we investigated in this paper), fixed and assigned according to some other probability distribution (see the discussion in Sec.~\ref{sec:discussion-networks}), or assigned in a time-dependent way (see the discussion in Sec.~\ref{sec:discussion-opinion}).

In network science, node weights have been studied far less than edge weights, and even the term ``weighted network'' usually refers specifically to edge-weighted networks by default. 
For example, it is very common to study centralities in edge-weighted networks \cite{opsahl2010}, but studies of centralities in node-weighted networks (e.g., see Refs.~\cite{heitzig2012, singh2020}) are much less common.
Heitzig et al. \cite{heitzig2012} generalized common network statistics to node-weighted networks and used node weights to represent the ``sizes'' of the nodes of a network.
They used their framework to study brain networks with node weights that encode the areas of regions of interest, international trade networks with node weights that encode the gross domestic products (GDPs) of countries, and climate networks with node weights that encode areas in a regular grid on the Earth's surface.
Singh et al.~\cite{singh2020} developed centrality measures that incorporate both edge weights and node weights and used them to study service-coverage problems and the spread of contagions.
These studies demonstrate the usefulness of node weights for incorporating salient information in network analysis in a variety of applications.

In our node-weighted BCM, we are interested in determining which nodes of a network are (in some sense) more influential than others and thereby exert larger effects on steady-state opinion-cluster profiles. 
Recently, Brooks and Porter \cite{brooks2020} quantified the influence of media nodes in a BCM by examining how their ideologies influence other nodes of a network.
An interesting area of future work is to develop ways to quantify the influence of specific nodes in models of opinion dynamics with node weights.
For example, can one determine which weighted nodes to seed with extreme opinions to best spread such opinions? Are there nodes that make it particularly easy for communities to reach consensus and remain
connected in a steady-state effective-receptivity network $G_{\mathrm{eff}}(T)$?
One can adapt the node weights in our BCM to examine a variety of sociological scenarios in which nodes have heterogeneous activity levels or interaction frequencies. More generally, our model illustrates the importance of incorporating node weights into network analysis, and we encourage researchers to spend more time studying the effects of node weights on network structure and dynamics. 


\begin{acknowledgements}

We thank Andrea Bertozzi, Jacob Foster, Jerry Luo, Deanna Needell, and the participants of UCLA’s Networks Journal Club for helpful comments and discussions. We also thank the two anonymous referees for helpful comments. We acknowledge financial support from the National Science Foundation (grant number 1922952) through the Algorithms for Threat Detection (ATD) program. GJL was also supported by NSF grant number 1829071.

\end{acknowledgements}




\providecommand{\noopsort}[1]{}\providecommand{\singleletter}[1]{#1}%
%


\end{document}